\documentclass[aps,twocolumn,prd,showpacs,showkeys,preprintnumbers,superscriptaddress,bibnotes,floatfix,longbibliography,notitlepage,nofootinbib]{revtex4-2}

\pdfoutput=1
\usepackage{amsmath}
\usepackage{amsfonts}
\usepackage{amssymb}
\usepackage{mathrsfs}
\usepackage{graphicx}
\usepackage{color}
\usepackage{bbding}
\usepackage{tabularx}
\usepackage[usenames,dvipsnames,table]{xcolor}
\usepackage[normalem]{ulem}
\usepackage{makecell}
\usepackage{siunitx}
\usepackage{enumerate}
\usepackage{multirow}
\usepackage{makecell}
\usepackage{upgreek}
\usepackage{blindtext}
\definecolor{darkred}{rgb}{0.5,0,0}
\definecolor{darkblue}{rgb}{0,0,0.5}
\definecolor{firebrick}{rgb}{0.75,0.125,0.125}
\definecolor{darkgreen}{rgb}{0,0.5,0}
\usepackage[colorlinks=true,linkcolor=firebrick,citecolor=darkgreen,urlcolor=darkblue]{hyperref} 
\usepackage[capitalise]{cleveref}

\newcommand{\ie}{{\it i.e.}}

\newcommand{\eg}{{\it e.g.}}

\newcommand{\eq}{Eq.}

\newcommand{\fig}{Fig.}

\newcommand{\Refe}{Ref.}
\newcommand{\Refes}{Refs.}

\newcolumntype{L}{>{\centering\arraybackslash}m{5cm}}
\newcommand{\equ}[1]{\eq~(\ref{equ:#1})}
\newcommand{\figu}[1]{\fig~\ref{fig:#1}}
\newcommand{\orcid}[1]{\href{https://orcid.org/#1}{\includegraphics[width=10pt]{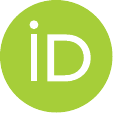}}}

\graphicspath{{figures/}{figures_appendix/}}

\begin{document}


\title{Joint measurement of the ultra-high-energy neutrino spectrum and cross section}

\author{V\'ictor B.~Valera
\orcid{0000-0002-0532-5766}}
\email{vvalera@nbi.ku.dk}
\affiliation{Niels Bohr International Academy, Niels Bohr Institute,\\University of Copenhagen, DK-2100 Copenhagen, Denmark}

\author{Mauricio Bustamante
\orcid{0000-0001-6923-0865}}
\email{mbustamante@nbi.ku.dk}
\affiliation{Niels Bohr International Academy, Niels Bohr Institute,\\University of Copenhagen, DK-2100 Copenhagen, Denmark}

\author{Olga Mena
\orcid{0000-0001-5225-975X}}
\email{omena@ific.uv.es}
\affiliation{Instituto de F\'isica Corpuscular (IFIC), Universidad de Valencia-CSIC, 46071, Valencia, Spain}

\date{August 14, 2023}

\begin{abstract}
 Soon, a new generation of neutrino telescopes, presently under planning, will target the discovery of ultra-high-energy (UHE) neutrinos of cosmic origin, with energies higher than 100~PeV, that promise unique insight into astrophysics and particle physics.  Yet, predictions of the UHE neutrino flux and interaction cross section---whose measurement is co-dependent---are laden with significant uncertainty that, if unaddressed, could misrepresent the capabilities to measure one or the other.  To address this, we advocate for the joint measurement of the UHE neutrino spectrum and neutrino-nucleon cross section, including of their energy dependence, without assuming prior knowledge of either.  We illustrate our methods by adopting empirical parametrizations of the neutrino spectrum, in forecasts geared to the planned radio array of the IceCube-Gen2 neutrino telescope.  We warn against using simple parametrizations---a simple power law or one augmented with an exponential cut-off---that might fail to capture features of the spectrum that are commonplace in the predictions.  We argue instead for the use of flexible parametrizations---a piecewise power law or an interpolating polynomial---that ensure  accuracy.  We report loose design targets for the detector energy and angular resolution that are compatible with those under present consideration.
\end{abstract}

\maketitle


\section{Introduction}
\label{sec:introduction}

For over fifty years~\cite{Berezinsky:1969erk}, ultra-high-energy (UHE) neutrinos, with EeV-scale energies (1 EeV $\equiv 10^{18}$~eV)---have evaded detection.  At long last, this might change in the next decade, thanks to new, larger neutrino telescopes.  Yet, because much is unknown about them, planning for their discovery has unavoidably involved making informed assumptions; notably, regarding the size of the UHE neutrino flux and interaction cross section, and their dependence on neutrino energy.  Below, we show that these assumptions can be abandoned---and should be abandoned---if we are to tap into the unbiased joint sensitivity to astrophysics and particle physics that the next generation of UHE neutrino telescopes will offer.

Ultra-high-energy neutrinos are produced in the interaction of ultra-high-energy cosmic rays (UHECRs) with radiation or matter, either inside extragalactic cosmic-ray accelerators, or during the long propagation of UHECRs to Earth~\cite{Ackermann:2022rqc}.  They are the most energetic neutrinos expected, 10--100 times more so than the TeV--PeV neutrinos discovered by the IceCube neutrino telescope that are the most energetic ones detected to date~\cite{IceCube:2020wum, IceCube:2021uhz}.  They carry insight into long-standing open questions in astrophysics~\cite{Ackermann:2019ows}---what are the sources of UHECRs---and fundamental physics~\cite{Ackermann:2019cxh}---how do neutrinos, and particle physics broadly, behave at the highest energies.  

The existence of UHE neutrinos is all but guaranteed, since it relies on the existence of UHECRs and cosmological radiation fields, both of which are observed.  However, they are rare; so much so that, so far, they remain undiscovered in spite of the fact that the neutrino-nucleon cross section, which fixes the probability of detecting UHE neutrinos, is expected to be larger at ultra-high energies~\cite{CTEQ:1993hwr, Gandhi:1995tf, Gandhi:1998ri, Conrad:1997ne, Formaggio:2012cpf, Connolly:2011vc, Cooper-Sarkar:2011jtt}.  Searches for UHE neutrinos have placed increasingly tighter upper limits on their diffuse flux~\cite{IceCube:2018fhm, Anker:2019rzo, ARA:2019wcf, PierreAuger:2019ens}, which means that existing neutrino telescopes are likely too small to discover them.  In parallel, predictions of the UHE neutrino flux have grown in sophistication, but still vary significantly in size and shape (see, \eg, \Refes~\cite{Aloisio:2009sj, Kotera:2010yn, Ahlers:2012rz, Fang:2013vla, Padovani:2015mba, Fang:2017zjf, Romero-Wolf:2017xqe, AlvesBatista:2018zui, Heinze:2019jou, Muzio:2019leu, Rodrigues:2020pli, Anker:2020lre, Muzio:2021zud} and Fig.~2 in \Refe~\cite{Valera:2022wmu}), because they rely on properties of UHECRs and their sources, which are only known uncertainly~\cite{Anchordoqui:2018qom, AlvesBatista:2019tlv}. 

In view of this, in the next 10--20 years a new generation of UHE neutrino telescopes, presently under planning and construction~\cite{Ackermann:2022rqc, Abraham:2022jse}, will target the discovery of UHE neutrinos even if their flux is tiny.  They adopt new detection strategies, or extend proven ones, to monitor detector volumes larger than present-day  telescopes---Baikal-GVD~\cite{Allakhverdyan:2021vkk} (following Baikal NT-200~\cite{Balkanov:1999ys}), KM3NeT~\cite{KM3Net:2016zxf} (following ANTARES~\cite{ANTARES:2011hfw}), and IceCube~\cite{IceCube:2013low}---which use in-ice and in-water optical Cherenkov detection.  Techniques under consideration include the detection of neutrino-initiated showers via their radio emission in ice, in IceCube-Gen2~\cite{IceCube-Gen2:2020qha}, RNO-G~\cite{RNO-G:2020rmc}, and RET-N~\cite{Prohira:2019glh} (following ARA~\cite{Allison:2011wk} and ARIANNA~\cite{ARIANNA:2019scz}), or in the atmosphere, in BEACON~\cite{Wissel:2020sec}, GRAND~\cite{GRAND:2018iaj}, PUEO~\cite{PUEO:2020bnn} (following ANITA), and TAROGE-M~\cite{TAROGE:2022soh}; via their Cherenkov emission in the atmosphere, like Ashra NTA~\cite{Sasaki:2014mwa}, CTA~\cite{Fiorillo:2020xst}, EUSO-SPB~\cite{Adams:2017fjh}, SKA~\cite{James:2017pvr}, TAx4~\cite{Kido:2019enj}, Trinity~\cite{Otte:2018uxj}, and, from space, POEMMA~\cite{POEMMA:2020ykm}; and via their particle showers at ground level, like AugerPrime~\cite{PierreAuger:2016qzd}.

\begin{figure*}[t]
 \centering
 \includegraphics[width=\textwidth]{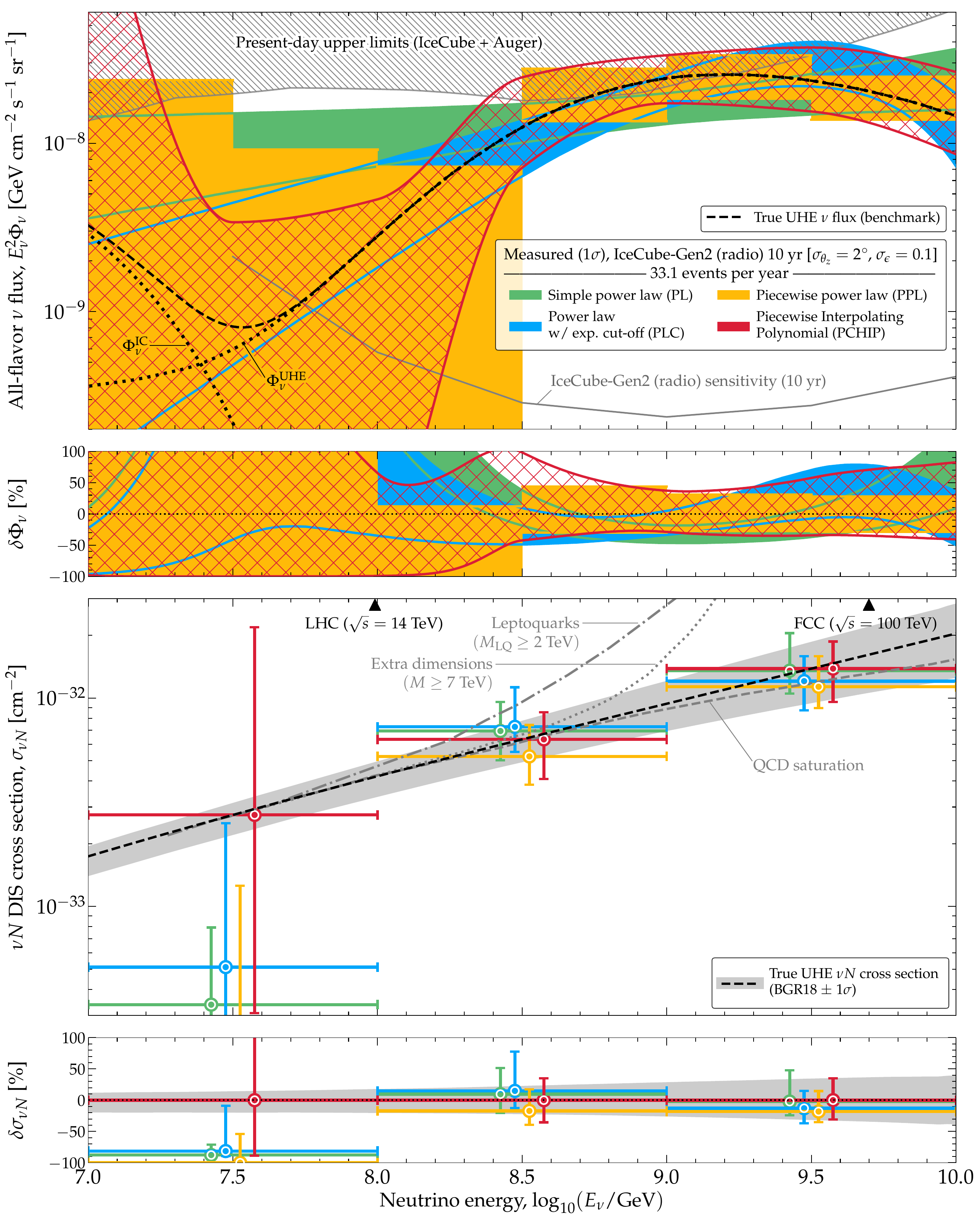}
 \caption{\textbf{\textit{Projected joint measurement of the ultra-high-energy (UHE) neutrino flux and neutrino-nucleon deep-inelastic scattering cross section in the radio array of IceCube-Gen2.}}  We show results for our benchmark UHE neutrino flux (Sec.~\ref{sec:nu_flux-benchmark}), computed using four fit models of the neutrino spectrum: PL, PLC, PPL, and PCHIP (Sec.~\ref{sec:models-flux}).  \textit{Top two panels}:  UHE neutrino flux, and relative error between the measured and true flux, $\delta \Phi_\nu = (\Phi_\nu - \Phi_\nu^{\rm true})/\Phi_\nu^{\rm true}$.  Present-day upper limits on the flux are from IceCube~\cite{IceCube:2018fhm} and Auger~\cite{PierreAuger:2019ens}.  The projected sensitivity of IceCube-Gen2 is from \Refe~\cite{IceCube-Gen2:2021rkf}.  \textit{Bottom two panels:}  UHE neutrino-nucleon cross section, and relative error between the measured and true~\cite{Bertone:2018dse} cross section, $\delta \sigma_{\nu N} = (\sigma_{\nu N} - \sigma_{\nu N}^{\rm true})/\sigma_{\nu N}^{\rm true}$.  Central values are offset horizontally for clarity.  See Sec.~\ref{sec:results} and \figu{alternative_models} for details.}
 \label{fig:main_result}
\end{figure*}

In preparation, previous works have forecast the potential of upcoming detectors to discover the diffuse flux of UHE neutrinos~\cite{Valera:2022wmu} (and also UHE neutrino point sources~\cite{Fang:2016hop, Fiorillo:2022ijt}), and to measure, for the first time, the UHE neutrino-nucleon cross section~\cite{Denton:2020jft, Valera:2022ylt, Esteban:2022uuw}.  The forecasts are encouraging.  However, they rely on key simplifying assumptions: they fix either the UHE neutrino flux---its normalization or energy spectrum---the neutrino-nucleon cross section---its size or energy dependence---or both.  Adopting these assumptions does not invalidate the conclusions reached by these works, though they may have cast them in an overly optimistic light, since these quantities are ultimately unknown and must be measured.  More importantly, they may have unwittingly downplayed the capability of upcoming neutrino telescopes to make joint measurements of the UHE neutrino flux and cross section, including of their energy dependence, unencumbered by assumptions on one or the other, and with little theoretical bias.

\textit{An earnest analysis of UHE neutrinos, one that conveys the true capabilities of upcoming neutrino telescopes, should strive to jointly measure the neutrino spectrum and the energy dependence of the neutrino-nucleon cross section.  For the first time, we provide forecasts of such an analysis.}  We show that this should be feasible under realistic experimental setups, so as to motivate upcoming experimental endeavors to embrace their full potential.

Figure~\ref{fig:main_result} shows a condensed view of our main results; we defer details to later.  We gear our results to UHE neutrino radio-detection in IceCube-Gen2 because it is one of the largest planned telescopes; we model it in state-of-the-art detail~\cite{Valera:2022ylt}.  Based on plausible detector performance and energy and angular resolution, \figu{main_result} shows that it should be possible to jointly measure the UHE neutrino spectrum and the energy dependence of the cross section, both with a relative uncertainty of tens of percent, enough to reconstruct the spectrum closely---which can provide astrophysical insight---and to identify potential deviations of the cross section from its standard prediction---which can provide particle-physics insight.

There are, however, three caveats; we address them in detail later.  First, such joint measurement requires a detection rate of at least a handful of UHE neutrinos per year; ideally, of a few tens.  However, this is not unique to our work; it is also true for previous forecasts of flux discovery~\cite{Valera:2022wmu} and cross-section measurement~\cite{Denton:2020jft, Valera:2022ylt, Esteban:2022uuw}.  Second, \figu{main_result} shows that measurements below 100~PeV are poor; this is because there UHE neutrino telescopes become less efficient.  This showcases the need for complementarity between UHE neutrino telescopes and telescopes sensitive to PeV-scale neutrinos~\cite{IceCube-Gen2:2020qha, Ackermann:2022rqc, vanSanten:2022wss}.  Third, for our measurements to be accurate, \ie, centered on the true values of the flux and cross section, it is critical when contrasting test {\it vs.}~true UHE neutrino energy spectra to use a parametrization of the test spectra that is flexible enough to capture the features of the true spectrum.  Otherwise, precision may be high but accuracy low.

The rest of this paper is organized as follows.  Section~\ref{sec:synopsis} presents the synopsis, context, and tenet of our work.  Section~\ref{sec:nu_flux} overviews general features of the UHE neutrino flux and introduces our benchmark flux model. Section~\ref{sec:nu_detection} describes the method we use to estimate event rates.  Section~\ref{sec:models} introduces the models of the UHE neutrino flux and cross section that we adopt in our fits to simulated observations. Section~\ref{sec:methods_stat} introduces our statistical methods.  Section~\ref{sec:results} shows results using different analysis choices.  Section~\ref{sec:summary} summarizes.


\section{Synopsis, context, and tenet}
\label{sec:synopsis}

\subsection{Measuring the neutrino flux and cross section}
\label{sec:synopsis-measuring}

Upon reaching the Earth, high-energy cosmic neutrinos travel from its surface, through its interior, to the detector, over distances of up to the diameter of the Earth.  Along the way, they interact with underground matter---predominantly via neutrino-nucleon ($\nu N$) deep inelastic scattering (DIS)~\cite{IceCube:2017roe, Bustamante:2017xuy, IceCube:2018pgc, IceCube:2020rnc}---and, as a result, their flux is attenuated.  Roughly, the attenuation factor is $e^{-\sigma_{\nu N} L}$, where $\sigma_{\nu N}$ is the $\nu N$ DIS cross section and $L$ is the distance traveled by the neutrinos inside Earth.  Thus, the flux of neutrinos that reach a detector from below, after traveling through thousands of kilometers inside the Earth, is more severely attenuated than the flux of neutrinos that reach it from above, after traveling through only a few kilometers of matter.  Because the $\nu N$ cross section grows with neutrino energy~\cite{CTEQ:1993hwr, Conrad:1997ne, Formaggio:2012cpf}, the attenuation is more severe the higher the neutrino energy.

The neutrino flux, $\Phi_\nu$, and the $\nu N$ cross section determine the number of neutrinos detected upon reaching the detector.  Roughly, this is $N_\nu \sim \Phi_\nu \sigma_{\nu N} e^{-\sigma_{\nu N} L}$.  This illustrates an essential aspect of our plight: for a given number of detected neutrinos, there is a degeneracy between the flux and cross section that is broken only by the exponential dampening term, as long as it is sufficiently different from unity.  It is, in fact, from this interplay that the measurement of the $\nu N$ cross section stems; we outline this below.  Yet, when the cross section is too high---at the highest energies---the attenuation nearly fully dampens the flux.  (We use the above expression only for illustration; later, in Sec.~\ref{sec:nu_detection}, we produce our results using detailed computations of neutrino propagation inside the Earth and detection.)

Below about 10~TeV---below the energies relevant to our work---the cross section is small enough for Earth to be largely transparent to neutrinos, regardless of their arrival directions to the detector, \ie, $e^{-\sigma_{\nu N} L} \approx 1$.  Between about 100~TeV and a few PeV---still below our energies of interest---the cross section grows to appreciably attenuate the flux of neutrinos that reach the detector from below, \ie, $e^{-\sigma_{\nu N} L} \lesssim 1$ from below and $e^{-\sigma_{\nu N} L} \approx 1$ from above.  Above 100~PeV---the energies relevant to our work---the attenuation is severe across most arrival directions, \ie, $e^{-\sigma_{\nu N} L} \ll 1$ from below $e^{-\sigma_{\nu N} L} \lesssim 1$ from above and from horizontal directions.  See Fig.~A2 in \Refe~\cite{Bustamante:2017xuy} for an illustration.  Later, we show how dwindling event rates are a challenge to UHE measurements.  

Our goal is to measure the UHE neutrino flux and $\nu N$ cross section---and to do it jointly.  For TeV--PeV neutrinos, these measurements are performed regularly using IceCube data, though seldom jointly (more on this below).  Given a sample of detected neutrinos, to measure the flux responsible for it, analyses undo the effect of in-Earth attenuation and, to measure the cross section, analyses compare the flux attenuation along different directions.  Below, we outline the methods used to measure the TeV--PeV neutrino flux and cross section.  At ultra-high energies, the methods are similar, but have important differences that motivate our analysis choices.


\subsection{Today: TeV--PeV measurements}

\textbf{\textit{Measuring the neutrino flux.---}}To infer the flux of TeV--PeV cosmic neutrinos by undoing the effect of in-Earth attenuation, analyses require knowledge of the $\nu N$ cross section, at least within its theoretical uncertainty, and an ansatz for the neutrino spectrum.  For the former, there are multiple predictions~\cite{Gandhi:1995tf, Gandhi:1998ri, Cooper-Sarkar:2007zsa, Gluck:2010rw, Connolly:2011vc, Cooper-Sarkar:2011jtt, Block:2014kza, Goncalves:2014woa, Arguelles:2015wba, Albacete:2015zra, Gauld:2016kpd, Ball:2017otu, Bertone:2018dse}.  For the latter, the nominal assumption, motivated by astrophysical considerations, is a power-law spectrum, $\Phi_\nu \propto E_\nu^{-\gamma}$, where $E_\nu$ is the neutrino energy.  The values of the flux normalization and the spectral index, $\gamma$, are allowed to float and are fixed in a fit to data that contrasts predicted event rates {\it vs.}~observed ones.  Numerous analyses using IceCube data have used this procedure to measure the diffuse flux of astrophysical and atmospheric neutrinos~\cite{IceCube:2013cdw, IceCube:2013low, IceCube:2013gge, IceCube:2014stg, IceCube:2014rwe, IceCube:2015mgt, IceCube:2015gsk, IceCube:2015qii, IceCube:2016umi, IceCube:2020acn, IceCube:2020wum, IceCube:2021uhz}. 

At present, there is no strong preference from data for alternative choices of the TeV--PeV neutrino spectrum---a broken power law, a power law with an exponential cut-off, or a log-parabola; see, \eg, \Refes~\cite{IceCube:2020wum, IceCube:2021uhz, Fiorillo:2022rft}.  Another, more flexible choice, trades the use of a single spectrum over the full energy range of the analysis for a piecewise spectrum that is $\propto E_\nu^{-2}$ inside multiple, relatively narrow energy bins, each one carrying its own independent, floating normalization.  The greater shape flexibility of a piecewise power-law spectrum grants the analysis sensitivity to narrow energy features that would otherwise be missed by using a rigid spectral shape.  We explore this choice ourselves later in our UHE forecasts (Sec.~\ref{sec:models-flux}).

\smallskip

\textbf{\textit{Measuring the $\nu N$ cross section.---}}The measurement of the $\nu N$ cross section stems from comparing the number of neutrinos of different energies detected along different arrival directions to the detector, which reflects how much the flux has been attenuated by propagating underground~\cite{Hooper:2002yq, Hussain:2006wg, Borriello:2007cs, Hussain:2007ba, Connolly:2011vc, Marfatia:2015hva}.  Based on this method, \Refes~\cite{IceCube:2017roe, Bustamante:2017xuy, IceCube:2018pgc, IceCube:2020rnc} used IceCube neutrinos to pioneer measurements of the TeV--PeV $\nu N$ DIS cross section.

However, most of these cross-section analyses made simplifying assumptions that restricted their purview.  Reference~\cite{IceCube:2017roe} measured only the normalization of the $\nu N$ cross section, while keeping its energy dependence fixed to a standard prediction~\cite{Cooper-Sarkar:2011jtt}.   Reference~\cite{IceCube:2020rnc} measured the energy dependence of the cross section, but did so by assuming a common neutrino power-law flux that spans the full energy range of the analysis.  Reference~\cite{IceCube:2018pgc} performed a related analysis, extracting the average inelasticity (see Sec.~\ref{sec:nu_detection-propagation}) of $\nu N$ DIS interactions.

\smallskip

\textbf{\textit{Joint measurements.---}}To the best of our knowledge, a joint measurement of the TeV--PeV neutrino flux and cross section, including of their energy dependence, has only been performed in full in \Refe~\cite{Bustamante:2017xuy}, using publicly available IceCube data~\cite{IceCube:2014stg, Kopper:2015vzf, IC4yrHESEURL, TalkKopperICRC2017}.  This analysis extracted the flux and cross section in separate energy bins, and provided the first measurement of the energy dependence of the cross section above 10~TeV.  Because the flux and cross section were measured independently in each bin, the results did not rely on assumptions of their behavior across a wide energy range, which could be either questionable or too constraining.  This conferred the analysis sensitivity to the potential presence of narrow features in the energy spectrum---such as bumps that could hint at the origin of the neutrinos~\cite{Fiorillo:2022rft}---or in the cross section---which could hint at deviations from standard predictions~\cite{Bustamante:2017xuy}.  (A later analysis \cite{IceCube:2020rnc} measured the cross section also in multiple energy bins, but assumed instead a common neutrino flux across them.)   


\subsection{This work: upcoming UHE measurements}

\textit{The upcoming access to ultra-high neutrino energies motivates making suitable analysis choices from the start.} 

\smallskip

The severe attenuation of the UHE neutrino flux inside the Earth shrinks not only the size of the available sample of detected events, but, more importantly, the range of neutrino directions from which we can compare the relative in-Earth attenuation to infer the $\nu N$ cross section.  Further, the regeneration of lower-energy neutrinos via neutral-current interactions and $\nu_\tau$ charged-current interactions~\cite{Garcia:2020jwr, Arguelles:2021twb}---relatively unimportant at lower energies---becomes important at ultra-high energies.  This renders the measurement of the neutrino flux and cross section at ultra-high energies especially susceptible to mismodeling their energy dependence.  

With this in mind, we adopt three guiding principles when forecasting measurements:
\begin{description}
 \item[Make joint measurements]
  We measure jointly the UHE neutrino flux and $\nu N$ cross section, and without fixing one when measuring the other.  In our statistical methods (Sec.~\ref{sec:methods_stat}), we adopt broad and uninformative priors on the flux and cross-section model parameters to avoid  introducing bias.
 \item[Measure the energy-dependent $\sigma_{\nu N}$]
  To be sensitive to the potential presence of narrow features in the $\nu N$ cross section or changes in its growth rate with energy---\eg, coming from extra dimensions~\cite{Jain:2000pu}, leptoquarks~\cite{Becirevic:2018uab}, or QCD saturation~\cite{Arguelles:2015wba}---we measure the cross section in multiple energy bins; see \figu{main_result}.
 \item[Measure the neutrino spectrum]
  Given the large variety in the shapes of competing predictions of the UHE neutrino spectrum (Sec.~\ref{sec:nu_flux}), we favor flexible parametrizations when attempting to measure it from data.  Later (Sec.~\ref{sec:results}), we show that adopting too rigid a parametrization leads to poor accuracy in the measurements; see \figu{main_result}.
\end{description}

We build an analysis based on these principles, gearing our forecasts to the radio-detection of UHE neutrinos in  IceCube-Gen2~\cite{IceCube-Gen2:2020qha}.  Our forecasts are based on methods introduced in \Refe~\cite{Valera:2022ylt}, and validated in \Refes~\cite{Fiorillo:2022ijt, Valera:2022wmu, Fiorillo:2023clw}.

Our goal is not to provide an exhaustive exploration of the capability to measure the UHE neutrino flux and cross section; for that, see \Refes~\cite{Valera:2022ylt, Valera:2022wmu}.  It is to bring attention to the need to measure them jointly, and to make suitable analysis choices when doing so.  Accordingly, later we produce forecasts assuming a relatively large benchmark UHE neutrino flux, based off of \Refe~\cite{Anker:2020lre}, that yields about 33 detected events per year (Sec.~\ref{sec:nu_flux}), which allows us to illustrate clearly our methods; see \figu{main_result}.

In reality, the rate of detected events may be appreciably lower.  In that case, we might be tempted to adopt simpler analysis choices rather than follow our guiding principles above, which involve more  informative, albeit complex, choices.  Granted, if only a handful of events were detected over the full observation campaign, simpler analysis choices would likely be preferable---\eg, using a simple power-law spectrum and measuring the cross-section normalization but not its energy dependence.  

However, we show below that maintaining simpler analysis choices when the event rate is higher would be detrimental.  Adopting too rigid a parametrization of the neutrino spectrum might unknowingly incur not only in its misreconstruction but also in that of the cross section (Sec.~\ref{sec:results}).  Still, choosing flexible parametrizations, with more model parameters, entails reducing the precision with which each parameter is measured:  selecting for accuracy comes at the cost of sacrificing some precision.  We deem this trade-off tolerable.


\section{Ultra-high-energy neutrinos}
\label{sec:nu_flux}


\subsection{Overview}
\label{sec:nu_flux-overview}

Ultra-high-energy neutrinos, with energies in excess of 100~PeV, are expected to be produced in the interaction of UHECRs, of EeV-scale energies, with matter or radiation, \ie, in $pp$ interactions \cite{Kelner:2006tc} or $p\gamma$ interactions \cite{Mucke:1999yb, Hummer:2010vx, Morejon:2019pfu}, respectively.  The interactions may occur inside the UHECR sources, purportedly extragalactic cosmic-ray accelerators---in which case neutrinos are dubbed {\it source neutrinos}---or during the propagation of UHECRs after leaving the sources, upon their scattering on the cosmic microwave background and the extragalactic background light---in which case they are dubbed {\it cosmogenic neutrinos}.  These interactions produce high-energy charged pions that decay into high-energy neutrinos, \ie, $\pi^+ \to \mu^+ + \nu_\mu$, followed by $\mu^+ \to e^+ + \nu_e + \bar{\nu}_\mu$, and their charge-conjugated processes.  Each neutrino carries about $5\%$ of the energy of the parent proton.  Other production processes contribute, too, their importance growing with neutrino energy~\cite{Mucke:1999yb, Hummer:2010vx, Morejon:2019pfu}. 

Broadly stated, the energy spectrum of UHE neutrinos made in $pp$ interactions is a power law $\propto E_\nu^{-\gamma}$ inherited from the power-law spectrum of the parent UHECR protons, with $2 \lesssim \gamma \lesssim 3$.  In contrast, the energy spectrum of UHE neutrinos made in $p\gamma$ interactions peaks at a characteristic energy set by the kinematic condition to produce the $\Delta(1232)$ resonance that decays into charged pions.  The shape of the spectrum, especially for source neutrinos, may be further affected by energy losses and gains of the parent cosmic rays and charged secondary particles~\cite{Winter:2013cla, Winter:2014tta, Bustamante:2020bxp}.   The normalization of the UHE neutrino flux depends on the properties of the UHECRs---their mass composition, spectrum, and maximum energy---and of their sources---mainly, their distribution in redshift.  The joint production of UHE source and cosmogenic neutrinos often considers both $pp$ and $p\gamma$ interactions; their predicted spectra are superpositions of a power law and a bump-like spectrum; \eg, \Refes~\cite{Fang:2017zjf, Muzio:2019leu, Muzio:2021zud}.   

Regarding the flavor composition of UHE neutrinos, \ie, the proportions of $\nu_e$, $\nu_\mu$, and $\nu_\tau$ in their total flux, the canonical expectation comes from the decay of high-energy pions (see above).  At the neutrino sources (S), the flavor composition from pion decay is $(f_{e, {\rm S}}, f_{\mu, {\rm S}}, f_{\tau, {\rm S}}) = (1/3, 2/3, 0)$, where $f_{\alpha, {\rm S}} \equiv (\Phi_{\nu_\alpha, {\rm S}} + \Phi_{\bar{\nu}_\alpha, {\rm S}})/\Phi_{\rm S}$ ($\alpha = e, \mu, \tau$), $\Phi_{\nu_\alpha, {\rm S}}$ is the flux of $\nu_\alpha$, $\Phi_{\bar{\nu}_\alpha, {\rm S}}$ is the flux of $\bar{\nu}_\alpha$, and $\Phi_{\rm S} \equiv \sum_{\alpha} ( \Phi_{\nu_\alpha, {\rm S}} + \Phi_{\bar{\nu}_\alpha, {\rm S}} )$ is the total neutrino flux.  Because of flavor mixing during propagation, the corresponding canonical flavor composition at Earth ($\oplus$), computed using the best-fit values of the neutrino mixing parameters~\cite{Esteban:2020cvm, NuFit_5.0}, is of near flavor equipartition: $( f_{e, \oplus}^\pi, f_{\mu, \oplus}^\pi, f_{\tau, \oplus}^\pi ) \approx ( 0.298, 0.359, 0.342 )$~\cite{Song:2020nfh} (see also \Refe~\cite{Bustamante:2015waa}).  By the years 2030--2040, for which we forecast, the uncertainties on the flavor composition at Earth are at the per-mille level~\cite{Song:2020nfh, Valera:2022ylt}, and we ignore them.  Other neutrino production processes or the physical conditions at the sources might affect the flavor composition, including in energy-dependent manners (see Fig.~5 in \Refe~\cite{Valera:2022ylt}), and so might new neutrino physics~\cite{Arguelles:2015dca, Bustamante:2015waa, Rasmussen:2017ert, Ackermann:2022rqc, Arguelles:2022tki}, though we do not consider those possibilities here.

In our forecasts of joint measurement of the UHE neutrino flux and cross section, we use the diffuse neutrino flux, made up of the contributions of all neutrino sources distributed across all redshifts, and which we take to be precisely isotropic.  It includes, in principle, contributions of source neutrinos and cosmogenic neutrinos; however, we make no attempt to tell these contributions apart.  The diffuse flux affords the largest expected rates of detected events.  In addition, the diffuse flux provides neutrinos from all arrival directions---which we need to make the measurements (Sec.~\ref{sec:synopsis-measuring})---unlike UHE neutrinos from point sources~\cite{Fiorillo:2022ijt}, which are confined to come from the directions of a limited number of sources.    

Presently, because the identity of the UHECR sources and the UHECR production processes are unknown, there is a large spread in the UHE neutrino flux predictions, in shape and size.  Figure~2 in \Refe~\cite{Valera:2022wmu} illustrates the current status of the flux predictions.

Regarding the neutrino spectrum, differences between competing predictions stem from different choices of the parent proton spectrum, target photon spectrum inside sources, and source properties such as magnetic field intensity.  For instance, models of neutrino production via $p\gamma$ interactions predict spectra that peak at $10^7$~GeV~\cite{Muzio:2021zud}, $10^8$~GeV~\cite{Rodrigues:2020pli}, or $10^9$~GeV~\cite{Anker:2019rzo}, with varying width of the spectrum around the peak value.  The spectrum might even have multiple peaks~\cite{Muzio:2021zud}.

Regarding the flux normalization, differences between competing predictions stem from the large uncertainty in the properties of UHECRs and their sources.  Heavier UHECR mass composition, steeper energy spectrum, and lower maximum energy---such as those favored by Pierre Auger Observatory (Auger) observations~\cite{PierreAuger:2020qqz, PierreAuger:2020kuy, PierreAuger:2022atd}---lead to a lower neutrino flux.  The converse---favored by Telescope Array (TA) observations~\cite{TelescopeArray:2018bya, TelescopeArray:2020bfv}---holds, too.  For details, see \eg, \Refes~\cite{Kotera:2010yn, Baerwald:2014zga, Romero-Wolf:2017xqe, AlvesBatista:2018zui, Heinze:2019jou}.  As a result, flux predictions range from being as high as to saturate the present-day upper limits~\cite{IceCube:2018fhm, PierreAuger:2019ens} (see \figu{main_result}) and as low as to be practically undetectable~\cite{Aloisio:2009sj, Ahlers:2012rz, Valera:2022wmu}.

Given the breadth of predictions of the UHE neutrino flux, we build our analysis methods later so that they are able to handle their variety of shape and size.  We achieve this by adopting flexible parametrizations of the spectrum, which we describe later (Sec.~\ref{sec:models-flux}).


\subsection{Benchmark UHE neutrino flux}
\label{sec:nu_flux-benchmark}

To illustrate our methods of joint measurement of the UHE neutrino flux and cross section, we adopt a benchmark model of the UHE neutrino flux that is representative of the breadth of flux predictions outlined above.  Following \Refe~\cite{Valera:2022wmu}, our benchmark flux is the sum of two flux components, shown in \figu{main_result}: an UHE flux prediction, $\Phi_{\nu_\alpha}^{\rm UHE}$, plus the UHE tail of the IceCube flux, $\Phi_{\nu_\alpha}^{\rm IC}$, \ie, $\Phi_{\nu_\alpha} = \Phi_{\nu_\alpha}^{\rm UHE} + \Phi_{\nu_\alpha}^{\rm IC}$.  When computing event rates due to our benchmark flux (Sec.~\ref{sec:event_rates}), we use the flux of each neutrino species separately, and assume equal proportions of neutrinos and anti-neutrinos of each flavor.  We describe each flux component below.

For the first component of our benchmark flux, $\Phi_{\nu_\alpha}^{\rm UHE}$, we adopt the prediction of the cosmogenic neutrino flux from \Refe~\cite{Anker:2019rzo}, generated by fitting the UHECR flux generated by a population of nondescript UHECR sources to observations by TA; see also \Refe~\cite{Valera:2022ylt} for an overview.  Because TA favors a relatively light UHECR mass composition at the highest energies~\cite{TelescopeArray:2020bfv}, the resulting cosmogenic neutrino flux is high---as high as is presently allowed by upper limits~\cite{IceCube:2018fhm, PierreAuger:2019ens}.  The flavor composition evolves with neutrino energy, but stays close to the canonical expectation from pion decay; see Fig.~6 in \Refe~\cite{Valera:2022ylt}. (This flux component is flux model 4 in \Refes~\cite{Valera:2022ylt, Valera:2022wmu}.)   This flux component is the dominant one: it yields about 33.1 events per year in the radio array of IceCube-Gen2, within $10^7$--$10^{10}$~GeV, computed using the methods that we introduce in Sec.~\ref{sec:event_rates}; see Table~I of \Refe~\cite{Valera:2022wmu} for detailed event rates. 

For the second component of our benchmark flux, $\Phi_{\nu_\alpha}^{\rm IC}$, we adopt an extrapolation to ultra-high energies of the TeV--PeV astrophysical neutrino power-law flux inferred by IceCube using 9.5~years of through-going $\nu_\mu$~\cite{IceCube:2021uhz}.  We augment it with an exponential cut-off at $E_{\nu, {\rm IC}} = 10^8$~GeV to explore the likely scenario where $\Phi_{\nu_\alpha}^{\rm IC}$ dies off at ultra-high energies and $\Phi_{\nu_\alpha}^{\rm UHE}$ takes over.  Thus, the second flux component is $\Phi_{\nu_\alpha}^{\rm IC} = (f_{\alpha, \oplus}^\pi / f_{\mu, \oplus}^\pi) (\Phi_{\nu_\mu+\bar{\nu}_\mu}/2) (E_\nu / 100~{\rm TeV})^{-\gamma_{\rm IC}} e^{-E_\nu/E_{\nu, {\rm IC}}}$, where $\Phi_{\nu_\mu+\bar{\nu}_\mu} = 1.44 \times 10^{-18}$~GeV$^{-1}$~cm$^{-2}$~s$^{-1}$~sr$^{-1}$ is the best-fit normalization of the $\nu_\mu + \bar{\nu}_\mu$ flux (dividing it by 2 splits it evenly between $\nu_\mu$ and $\bar{\nu}_\mu$) and $\gamma_{\rm IC} = 2.37$ is the best-fit spectral index, both from \Refe~\cite{IceCube:2021uhz}.  The prefactor $(f_{\alpha, \oplus}^\pi / f_{\mu, \oplus}^\pi)$ converts the flux of $\nu_\mu$ into the flux of $\nu_\alpha$, using the flavor ratios from Sec.~\ref{sec:nu_flux-overview}.  By itself, the second flux component yields a mean rate of only about $0.35$~events per year (see Table~I of \Refe~\cite{Valera:2022wmu}).  Regardless, they should not be ignored because they are concentrated in the low end of the IceCube-Gen2 energy range, whereas events from the first flux component are concentrated at higher energies.  (The background of atmospheric muons that we consider (Sec.~\ref{sec:bg_rates}) is similarly concentrated at low energies; see \figu{event_rate}.)

(Basing instead the second component of our benchmark flux on the IceCube neutrino flux inferred from 7.5~years of High Energy Starting Events (HESE)---a steeper power law with spectral index of around 2.87---would have contributed a negligible rate of 0.22 events in 10 years; see Table I and Fig.~12 in \Refe~\cite{Valera:2022wmu}.)


\section{UHE neutrino detection}
\label{sec:nu_detection}


\subsection{Neutrino propagation inside Earth}
\label{sec:nu_detection-propagation}

After neutrinos arrive at Earth, they propagate underground, from the surface of the Earth to the detector, IceCube-Gen2, located in South Pole, over distances of up to the diameter of the Earth.  Along the way, they interact with underground Earth matter, which modifies the flux that reaches the detector depending on neutrino energy, direction, and flavor~\cite{Gandhi:1998ri, Connolly:2011vc, IceCube:2017roe, Bustamante:2017xuy, IceCube:2018pgc, IceCube:2020rnc, Garcia:2020jwr, Denton:2020jft, Valera:2022ylt, Valera:2022wmu, Esteban:2022uuw}.  In our analysis, we account for these effects using state-of-the-art calculations of the neutrino-matter cross sections and of neutrino propagation inside Earth.  We present an overview below, and defer to \Refes~\cite{Garcia:2020jwr, Valera:2022ylt, Valera:2022wmu} for details.

At the energies of interest, the neutrino interaction on matter is typically $\nu N$ DIS.  In a scattering event, the interacting neutrino scatters off of one of the partons---a quark or a gluon---of a nucleon, $N$---a proton or a neutron---and breaks it up into final-state hadrons, $X$.  The interaction is either neutral-current (NC), \ie, $\nu_\alpha + N \to \nu_\alpha + X$, where the final-state neutrino has lower energy than the incoming neutrino, or charged-current (CC), \ie, $\nu_\alpha + N \to l_\alpha + X$, where $l_\alpha$ is a charged-current lepton of flavor $\alpha$.  Anti-neutrinos undergo the same processes, charge-conjugated; at ultra-high energies, the $\bar{\nu} N$ and $\nu N$ DIS cross sections are very similar.

The final-state hadrons receive a fraction $0 \leq y \leq 1$, the inelasticity, of the energy of the incoming neutrino, while the final-state lepton receives the remaining fraction $1-y$.  The inelasticity distributions, $d\sigma_{\nu N}^{\rm NC}/dy$ and $d\sigma_{\nu N}^{\rm CC}/dy$, where $\sigma_{\nu N}^{\rm NC}$ and $\sigma_{\nu N}^{\rm CC}$ are the NC and CC cross sections, are energy-dependent and relatively broad; see, \eg, Fig.~4 in \Refe~\cite{Valera:2022ylt}.  At ultra-high energies, the average value of the inelasticity is $\langle y \rangle \approx 0.25$~\cite{Gandhi:1995tf}; however, we compute neutrino interactions during propagation using the $y$ distributions, not their averages.

Because the $\nu N$ DIS cross sections grow with energy, roughly $\propto E_\nu^{0.36}$ at ultra-high energies~\cite{Gandhi:1998ri}, the $\nu N$ interactions of UHE neutrinos significantly modify the flux that reaches the detector.  While NC interactions shift the neutrino flux to lower energies, CC interactions deplete it.  Broadly stated, the net effect is an exponential attenuation of the neutrino flux, more prominent the higher the energy and the longer the distance traveled inside Earth (Sec.~\ref{sec:synopsis-measuring}); see, \eg, Figs.~A1 and A2 of \Refe~\cite{Bustamante:2017xuy}, Fig.~4.1 of \Refe~\cite{Garcia:2020jwr}, and Fig.~9 of \Refe~\cite{Valera:2022ylt}.  As a result, at ultra-high energies, the flux of upgoing neutrinos, \ie, those coming from $\theta_z \gtrsim 90^\circ$ ($\theta_z$ is the zenith angle measured from the South Pole) is nearly fully attenuated when it reaches the detector.  In contrast, the flux of neutrinos arriving from downgoing ($\theta_z \gtrsim 90^\circ$) and horizontal ($\theta_z \approx 90^\circ$) directions is appreciably, though not fully attenuated.  It is from these directions that UHE neutrinos may be discovered~\cite{Fiorillo:2022ijt, Valera:2022wmu} and their cross section measured~\cite{Hooper:2002yq, Connolly:2011vc, Denton:2020jft, Valera:2022ylt, Esteban:2022uuw}.  Our joint measurement of the UHE neutrino energy spectrum and cross section stems from neutrinos from these directions (Sec.~\ref{sec:results}).

In our work, we propagate UHE neutrinos inside the Earth using the same procedure as in \Refes~\cite{Valera:2022ylt, Fiorillo:2022ijt, Valera:2022wmu}.  We use {\sc NuPropEarth}~\cite{Garcia:2020jwr, NuPropEarth}, a state-of-the-art Monte-Carlo neutrino propagation code.  For the underground matter density of the Earth, we use the Preliminary Reference Earth Model~\cite{Dziewonski:1981xy}, including variations in the chemical composition with depth.  For the UHE $\nu N$ DIS cross sections, {\sc NuPropEarth} adopts the recent BGR18 calculation~\cite{Bertone:2018dse} and accounts for sub-leading neutrino interactions, for the important Glashow resonance of $\bar{\nu}_e$ on atomic electrons, and for the regeneration of $\nu_\tau$ via repeated CC interactions; see \Refe~\cite{Garcia:2020jwr} for details and \Refe~\cite{Valera:2022ylt} for a summary.  We propagate the fluxes of $\nu_e$, $\bar{\nu}_e$, $\nu_\mu$, $\bar{\nu}_\mu$, $\nu_\tau$, and $\bar{\nu}_\tau$ separately.  Unlike the fluxes at the surface of the Earth, the fluxes that reach the detector, $\Phi_{\nu_\alpha}^{\rm det}$ and $\Phi_{\bar{\nu}_\alpha}^{\rm det}$, are no longer isotropic.  For each neutrino flavor, initial energy, and direction, we propagate $10^7$ neutrinos using {\sc NuPropEarth}, as in \Refes~\cite{Valera:2022ylt, Fiorillo:2020xst, Valera:2022wmu, Fiorillo:2023clw}.   For the geometry of the radio array of IceCube-Gen2, our detector of choice, we use the cylindrical volume model introduced in \Refe~\cite{Valera:2022ylt}.  


\subsection{Neutrino radio-detection in IceCube-Gen2} \label{sec:event_rates}

\begin{figure*}
 \centering
 \includegraphics[width=\textwidth]{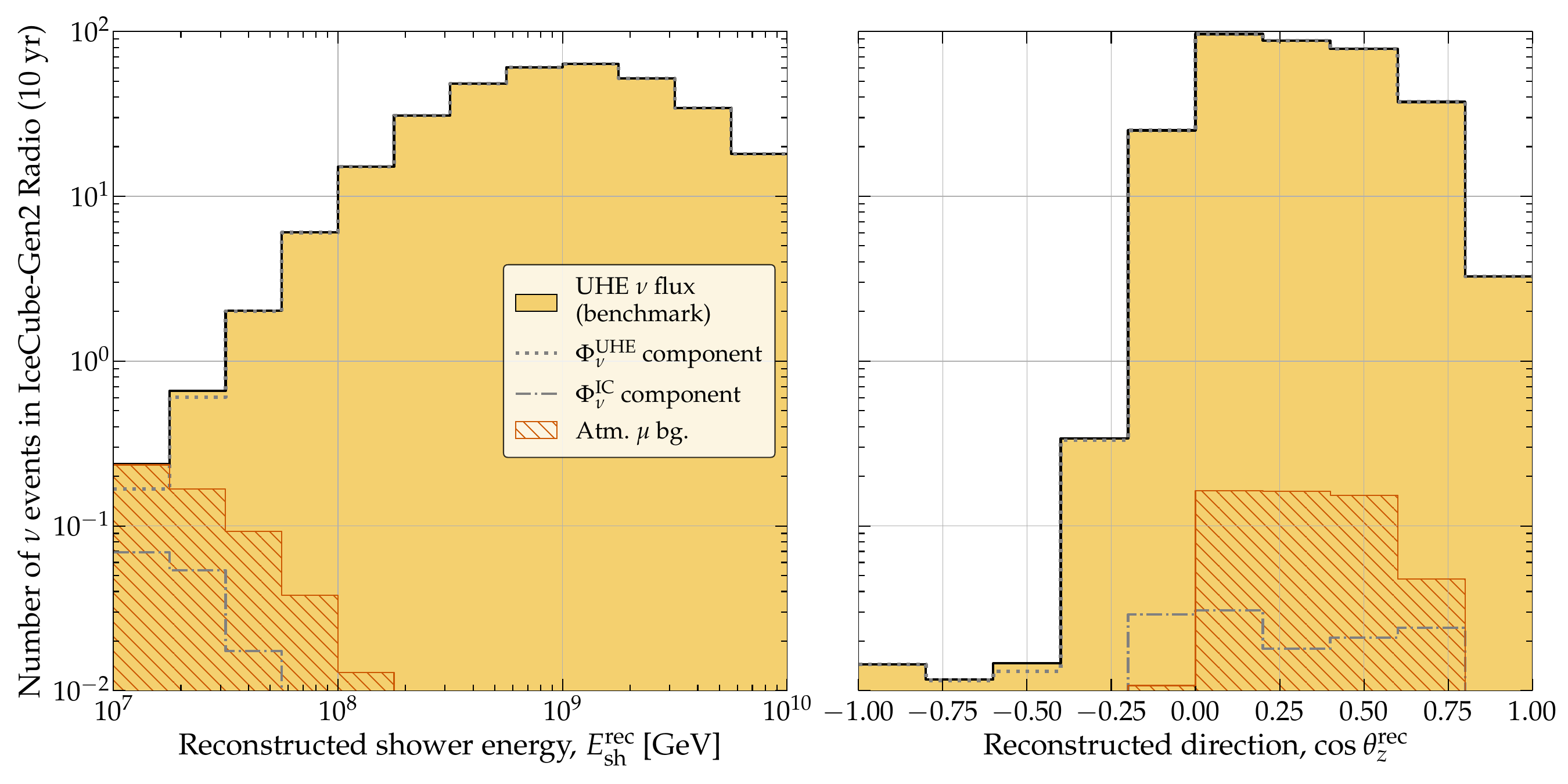}
 \caption{\textbf{\textit{Mean expected number of events detected in the radio array of IceCube-Gen2 induced by our benchmark neutrino flux model.}}  The benchmark neutrino flux is described in Sec.~\ref{sec:nu_flux-benchmark} and shown in \figu{main_result}. 
 It consists of a cosmogenic flux component, $\Phi_\nu^{\rm UHE}$, from \Refe~\cite{Anker:2019rzo}, and the UHE tail of the diffuse flux inferred from the IceCube 9.5-year $\nu_\mu$ analysis~\cite{IceCube:2021uhz}, augmented by an exponential cut-off at 100~PeV, $\Phi_\nu^{\rm IC}$.  Event rates are computed using our baseline choices for the detector energy resolution, $\sigma_{\epsilon} = 0.1$, and angular resolution, $\sigma_{\theta_z} = 2^\circ$.  The background of high-energy atmospheric muons is computed using the baseline prescription of the muon flux.  See Sec.~\ref{sec:event_rates} for the computation of neutrino-induced event rates and Sec~\ref{sec:bg_rates}, for that of event rates induced by atmospheric muons.  \textit{Left:} Events distributed in reconstructed energy. 
 \textit{Right:} Events distributed in reconstructed direction.}
 \label{fig:event_rate}
\end{figure*}

In the planned radio array of IceCube-Gen2~\cite{IceCube-Gen2:2020qha}, UHE neutrinos will be detected via their interaction with nucleons in the ice, which initiates a particle shower whose electromagnetic component emits coherent, impulsive radio signals. The radio emission, known as Askaryan radiation~\cite{Askaryan:1961pfb}, is due to the dipole formed between the shower axis and the excess of negative charges that accumulate on the front of the shower as it propagates.  IceCube-Gen2 will be instrumented with radio antennas buried in the ice, capable of detecting this signal.  From it, it will be possible to reconstruct the shower direction and energy, which are proxies for the direction and energy of the interacting neutrino.  To compute neutrino-initiated event rates, we use the methods and detector description introduced in \Refe~\cite{Valera:2022ylt}, to which we defer for details.  Below, we only outline them.

The relation between the neutrino energy, $E_\nu$, and the ensuing shower energy, $E_{\rm sh}$, depends on the neutrino flavor and on whether the interaction is NC or CC DIS.  For NC interactions of neutrinos of all flavors and CC interactions of $\nu_\mu$ and $\nu_\tau$, $E_\nu^{\rm NC} = E_{\rm sh}/y$, since only the final-state hadrons radiate~\cite{Garcia-Fernandez:2020dhb}. 
For CC interactions of $\nu_e$, $E_\nu^{\rm CC} = E_{\rm sh}$, since both the final-state charged electron and hadrons radiate.  Like during in-Earth propagation, the inelasticity, $y$, is distributed following the differential cross section (Sec.~\ref{sec:nu_detection-propagation}).  At ultra-high energies, the $y$ distributions of $\nu_\alpha$ and $\bar{\nu}_\alpha$ of all flavors are equal.

The detector response is expressed via simulated effective detector volumes, $V_{{\rm eff}, \nu_\alpha}^{\rm NC}$ and $V_{{\rm eff}, \nu_\alpha}^{\rm CC}$, for NC and CC interactions, respectively.  They are the same for $\nu_\alpha$ and $\bar{\nu}_\alpha$.  The effective volumes that we use were introduced in \Refe~\cite{Valera:2022ylt}, and used also in \Refes~\cite{Fiorillo:2022ijt, Valera:2022wmu, Fiorillo:2023clw}.  They are generated in simulations of neutrino interaction, and of the ensuing radio emission and propagation in ice, employing the same tools used by the IceCube-Gen2 Collaboration,  \textsc{NuRadioReco}~\cite{Glaser:2019cws} and \textsc{NuRadioMC}~\cite{Glaser:2019rxw}.  We adopt the same baseline detector configuration as in \Refes~\cite{IceCube-Gen2:2021rkf, Valera:2022ylt}, made up of a combination of shallow and deep radio stations.  The effective volumes vary with the shower direction, $\theta_z$, shower energy, $E_{\rm sh}$, and neutrino flavor; see Fig.~13 in \Refe~\cite{Valera:2022ylt} for an illustration.

The differential rate of events~\cite{Valera:2022ylt} initiated by $\nu_\alpha$ is
\begin{widetext}
 \begin{eqnarray}
  \label{equ:spectrum_true}
  \frac{d^2N_{\nu_\alpha}}{dE_{\rm sh} d\cos\theta_z}
  &=&
  2 \pi T n_t 
  \int_0^1 dy
  \left(
  \left.
  \frac{E_{\nu_\alpha}^{\rm {NC}}(E_{\rm sh}, y)}{E_{\rm sh}}
  V_{{\rm eff}, \nu_\alpha}^{\rm NC}(E_{\rm sh}, \cos\theta_z)
  \frac{d\sigma_{\nu_\alpha}^{\rm NC}(E_\nu, y)}{dy}
  \Phi^{\rm det}_{\nu_\alpha}(E_\nu,\cos\theta_z)
  \right\vert_{E_\nu = E_{\nu_\alpha}^{\rm NC}(E_{\rm sh}, y)}
  \right.
  \\
  &&
  \left.
  +~
  {\rm NC} \to {\rm CC}
  \right)
  \;,
  \nonumber
 \end{eqnarray}
\end{widetext}
where $T$ is the exposure time, $n_t$ is the number density of water molecules in ice, and $d\sigma_{\nu_\alpha {\rm w}}^{\rm NC}/dy$ is the $\nu N$ NC cross section on a water molecule, made up of 10 protons and 8 neutrons.  The event rate due to $\bar{\nu}_\alpha$ is the same as \equ{spectrum_true}, but changing $\Phi_{\nu_\alpha}^{\rm det} \to \Phi_{\bar{\nu}_\alpha}^{\rm det}$, $d\sigma_{\nu_\alpha {\rm w}}^{\rm NC}/dy \to d\sigma_{\bar{\nu}_\alpha {\rm w}}^{\rm NC}/dy$, and $d\sigma_{\nu_\alpha {\rm w}}^{\rm CC}/dy \to d\sigma_{\bar{\nu}_\alpha {\rm w}}^{\rm CC}/dy$. 

We account for the capability of the detector to reconstruct the shower energy and arrival direction by smearing the above differential event rate using the energy and angular resolution functions.  The detector reports the reconstructed shower energy, $E_{\rm sh}^{\rm rec}$, and direction, $\theta_z^{\rm rec}$, of the events.  For the energy resolution function, $\mathcal{R}_{E_{\rm sh}}(E_{\rm sh}^{\rm rec}, E_{{\rm sh}})$, we use a Gaussian function of $\log_{10} E_{\rm sh}^{\rm rec}$, centered at the true shower energy, $\log_{10} E_{\rm sh}$, with a width of $\sigma_{E_{\rm sh}} \equiv 10^{\sigma_\epsilon} E_{\rm sh}$, where   $\epsilon \equiv \log_{10}(E_{\rm sh}^{\rm rec}/E_{\rm sh})$; we choose $\sigma_\epsilon = 0.1$ for our baseline results~\cite{Anker:2019zcx, Aguilar:2021uzt, Gaswint:2021smu}. For the angular resolution function, $\mathcal{R}_{\theta_{z}}(\theta_{z}^{\rm rec}, \theta_z)$, we use a Gaussian function on the zenith angle with a width of $\sigma_{\theta_z}=2^\circ$, which approximates the expected point spread function of the radio array of IceCube-Gen2 of $\sigma_\Omega = 3^\circ$~\cite{Glaser:2019rxw, Gaswint:2021smu, ARIANNA:2021pzm, RNO-G:2021zfm}.  Reference~\cite{Valera:2022ylt} contains full definitions of the resolution functions.  We integrate over the real values of shower energy and direction to obtain the differential event rate in terms of measured quantities, 
\begin{eqnarray}
 \label{equ:spectrum_rec}
 \frac{d^2N_{\nu_\alpha}}
 {dE_{\rm sh}^{\rm rec} d\theta_{z}^{\rm rec}}
 &=&
 \int_{-1}^{+1} d\cos\theta_z
 \int_{0}^{\infty} dE_{\rm sh}
 \frac{d^2N_{\nu_\alpha}}{dE_{\rm sh} d\cos\theta_z} 
 \nonumber \\
 &&
 \times~
 \mathcal{R}_{E_{\rm sh}}(E_{\rm sh}^{\rm rec}, E_{{\rm sh}})~
 \mathcal{R}_{\theta_{z}}(\theta_{z}^{\rm rec}, \theta_z) \;,
\end{eqnarray}
and similarly for $\bar{\nu}_\alpha$.  In Sec.~\ref{sec:results-resolution} we comment on the effect that the energy and angular resolution have on our results.  We use the neutrino differential event rate due to $\nu_\alpha$ and $\bar{\nu}_\alpha$ of all flavors, obtained by adding their individual contributions,
\begin{equation}
 \label{equ:spectrum_tot}
 \frac{d^2N_{\nu}}
 {dE_{\rm sh}^{\rm rec} d\theta_{z}^{\rm rec}}
 =
 \sum_{\alpha=e,\mu,\tau}
 \left(
 \frac{d^2N_{\nu_\alpha}}
 {dE_{\rm sh}^{\rm rec} d\theta_{z}^{\rm rec}}
 +
  \frac{d^2N_{\bar{\nu}_\alpha}}
 {dE_{\rm sh}^{\rm rec} d\theta_{z}^{\rm rec}}
 \right) \;.
\end{equation}

Figure~\ref{fig:event_rate} shows the binned distribution of the mean number of events for our benchmark UHE neutrino flux (Sec.~\ref{sec:nu_flux-benchmark}), computed using \equ{spectrum_tot}.  The all-sky, energy-integrated event rate, within $E_{\rm sh}^{\rm rec} \in [10^7, 10^{10}]$~GeV, is about $\mathcal{N}_\nu = 33.1$ events per year.  The event distribution peaks around $10^9$~GeV because that is where the flux peaks and where the detector effective volume is larger.


\subsection{Atmospheric muon background}
\label{sec:bg_rates}

Separately, we compute the rate of events induced by the background of high-energy atmospheric muons, $d^2N_\mu / dE_{\rm sh}^{\rm rec}d\theta_z^{\rm rec}$.  To compute it, we follow the procedure introduced in \Refe~\cite{Valera:2022ylt}, using the same energy and angular resolution as for neutrino-initiated events above.  Like in \Refes~\cite{Valera:2022ylt, Fiorillo:2022ijt, Valera:2022wmu}, we adopt the calculation of the atmospheric muon background from \Refes~\cite{Garcia-Fernandez:2020dhb, Glaser:2021hfi, Hallmann:2021kqk}, based on the hadronic interaction model {\sc Sybill 2.3c}~\cite{Fedynitch:2018cbl}, accounting for the detector response, and mitigated by a surface array of cosmic-ray detectors that acts as veto.  Figure~15 in \Refe~\cite{Valera:2022ylt} shows the effect of the veto on event rates.  

Figure~\ref{fig:event_rate} shows the baseline prediction of the rate of muon-induced events (see also Figs.~15 and 16 in \Refe~\cite{Valera:2022ylt}, and Table I and Figs.~4, 5 in \Refe~\cite{Valera:2022wmu}).  The muon background is small---on average, we expect $\mathcal{N}_\mu = 0.54$ events in 10 years in the range $E_{\rm sh}^{\rm rec} \in [10^7, 10^{10}]$~GeV---is concentrated in the lower end of this range and is exclusively downgoing.  In spite of its small size, we do not ignore it because it can be relevant if the contribution of neutrino-initiated events is low---due to a low neutrino flux---and concentrated at low energies.  We make no attempt to distinguish the contribution of neutrino-initiated events from that muon-initiated events.  References~\cite{Fiorillo:2022ijt, Valera:2022wmu} explored the related issue of the influence of the muon background on the discovery of UHE neutrinos and the measurement of the UHE cross section.

(A potential second source of background events in the radio array of IceCube-Gen2 is due to the cores of particle showers initiated by cosmic rays in the atmosphere, which may penetrate the ice and continue to develop, producing Askaryan radiation alike that of neutrinos~\cite{DeKockere:2022bto}.  We do not include this background in our analysis because presently estimates of it are uncertain, though work is ongoing in improving and mitigating them.)


\section{Fit models of the UHE neutrino flux and cross section}
\label{sec:models}

Table~\ref{tab:parameters} shows a summary of the free parameters of the models of the UHE neutrino flux, $\nu N$ DIS cross section, and atmospheric muon background that we adopt, and our choice of priors for them.  We explore models that are relatively simple and that allow us to measure the flux and cross section in varying detail, though our focus is on flexible models that allow to measure the shape of the neutrino spectrum.  Below, we describe them; later, we use them in our statistical methods (Sec.~\ref{sec:methods_stat}) to produce our forecasts (Sec.~\ref{sec:results}).


\subsection{UHE neutrino flux}
\label{sec:models-flux}

We make forecasts using four competing models of the UHE neutrino  spectrum: a simple power law (PL), a power law with an exponential cut-off (PLC), a piecewise power law (PPL), and a Piecewise Cubic Hermite Interpolating Polynomial (PCHIP).  Each model has a different set of parameters, $\boldsymbol{f}_\Phi$, that describe the neutrino spectrum; we present them below.  In our forecasts (Sec.~\ref{sec:methods_stat}) we find values for $\boldsymbol{f}_\Phi$ via fits to projected observations in IceCube-Gen2.  The choice of flux model impacts the quality of the measurement of not only the neutrino spectrum, but also of the $\nu N$ cross section.

By construction, the fluxes of $\nu_\alpha$ and $\bar{\nu}_\alpha$ that we use are identical at the surface of the Earth, \ie, $\Phi_{\nu_\alpha} = \Phi_{\bar{\nu}_\alpha}$; hence the factor of 2 in the flux definitions, Eqs.~(\ref{equ:flux_pl})--(\ref{equ:flux_pchip}) below.  We assume a common spectral shape for neutrinos of all flavors.  The flavor composition at Earth, $f_{\alpha, \oplus}$, is fixed to the best-fit prediction for the year 2040~\cite{Song:2020nfh, Valera:2022wmu} (Sec.~\ref{sec:nu_flux}).  The above assumptions are common and reasonable simplifications.  In reality, each species of UHE neutrino could have a spectrum of its own; see, \eg, flux models 3--7 and 12 in Fig.~6 of \Refe~\cite{Valera:2022ylt}.  

\smallskip

\textbf{\textit{Power law (PL).---}}The UHE $\nu_\alpha$ spectrum is 
\begin{equation}
 \label{equ:flux_pl}
 \Phi_{\nu_\alpha}(E_\nu, \boldsymbol{f}_\Phi)
 =
 \frac{f_{\alpha,\oplus}^\pi}{2}
 \Phi_0
 \left(\frac{E_\nu}{10~{\rm PeV}}\right)^{-\gamma} \;.
\end{equation}
The free parameters are the flux normalization at 10~PeV, $\Phi_0$, and the spectral index, $\gamma$, \ie, $\boldsymbol{f}_\Phi \equiv (\Phi_0, \gamma)$.  In analyses of IceCube TeV--PeV neutrinos, a simple power law is the standard and often marginally preferred shape for the diffuse neutrino spectrum; see, \eg, \Refes~\cite{IceCube:2020wum, IceCube:2021uhz}.  (Yet, present-day uncertainties might conceal additional spectral features; see, \eg, \Refes~\cite{Bustamante:2020bxp, Fiorillo:2022rft}.)  For UHE neutrinos, most predictions of the spectrum have a structure richer than a simple power law; see Fig.~6 in \Refe~\cite{Valera:2022ylt} and Fig.~2 in \Refe~\cite{Valera:2022wmu}.  We use the PL model largely as a baseline against which to compare the other flux models.

\smallskip

\textbf{\textit{Power law with exponential cut-off (PLC).---}}The UHE $\nu_\alpha$ spectrum is
\begin{equation}
 \label{equ:flux_plc}
 \Phi_{\nu_\alpha}(E_\nu, \boldsymbol{f}_\Phi)
 =
 \frac{f_{\alpha,\oplus}^\pi}{2}
 \Phi_0
 \left(\frac{E_\nu}{10~{\rm PeV}}\right)^{-\gamma} 
 e^{-\frac{E_\nu}{E_{\rm cut}}} \;.
\end{equation}
The free parameters are the flux normalization at 10~PeV, $\Phi_0$, the spectral index, $\gamma$, and the cut-off energy, $E_{\rm cut}$, \ie, $\boldsymbol{f}_\Phi \equiv (\Phi_0, \gamma, E_{\rm cut})$.  In analyses of IceCube TeV--PeV neutrinos, present-day IceCube observations do not strongly disfavor a PLC diffuse neutrino spectrum with a cut-off in the multi-PeV range~\cite{IceCube:2021uhz}.  For UHE neutrinos, a large number of theoretical flux predictions resemble a PLC, albeit some of  only roughly; see Fig.~2 in \Refe~\cite{Valera:2022wmu}.

\smallskip

\textbf{\textit{Piecewise power law (PPL).---}}The UHE $\nu_\alpha$ spectrum is
\begin{eqnarray}
 \label{equ:flux_ppl}
 &&
 \Phi_{\nu_\alpha}(E_\nu, \boldsymbol{f}_\Phi)
 =
 \frac{f_{\alpha,\oplus}^\pi}{2}
 \left(\frac{E_\nu}{10~{\rm PeV}}\right)^{-2.0}
 \nonumber
 \\ &&
 \qquad\qquad
 \times
 \begin{cases}
     \Phi_1, & 10^7 \leq E_\nu/{\rm GeV} < 10^{7.5} \\
     \Phi_2, & 10^{7.5} \leq E_\nu/{\rm GeV} < 10^8 \\
     \Phi_3, & 10^8 \leq E_\nu/{\rm GeV} \leq 10^{8.5} \\
     \Phi_4, & 10^{8.5} \leq E_\nu/{\rm GeV} < 10^{9} \\
     \Phi_5, & 10^{9} \leq E_\nu/{\rm GeV} < 10^{9.5} \\
     \Phi_6, & 10^{9.5} \leq E_\nu/{\rm GeV} \leq 10^{10} \\
 \end{cases}
 \;.
\end{eqnarray}
The free parameters are the six flux normalization constants, $\Phi_i$ ($i = 1, \dots, 6$), one for each of the six half-decade energy bins between 10~PeV and 10~EeV, \ie, $\boldsymbol{f}_\Phi \equiv (\Phi_1, ..., \Phi_6)$.  In analyses of IceCube TeV--PeV neutrinos, this parametrization is often employed to infer the neutrino spectrum~\cite{IceCube:2021uhz}.  We use half-decade bins because this bin size is larger than our baseline choice for energy resolution of 10\% per energy decade (Sec.~\ref{sec:event_rates}).  Six bins is also about the minimum number needed to capture the main features of the neutrino spectrum.

\smallskip

\textbf{\textit{Piecewise Cubic Hermite Interpolating Polynomial (PCHIP).---}}A PCHIP is a shape-preserving interpolating function that uses monotonic cubic splines anchored at nodes with predefined positions~\cite{Fritsch:1984}.  In our case, we use it to interpolate the UHE neutrino energy spectrum based off of seven flux nodes $f_{\Phi, i} \equiv \log_{10}[E_{\nu, i}^2 \Phi_{\nu_\alpha}(E_{\nu, i})]$ ($i = 1, \dots, 7$), located at seven fixed values of the neutrino energy, $E_{\nu,i}$, equally spaced in logarithmic scale between $10^7$ and $10^{10}$~GeV.  For given values of  $\boldsymbol{f}_\Phi \equiv (f_{\Phi, 1}, \ldots, f_{\Phi, 7})$, we construct the PCHIP, $\varphi(E_\nu, \boldsymbol{f}_\Phi)$, and, with it, the flux of $\nu_\alpha$, as
\begin{equation}
 \label{equ:flux_pchip}
 \Phi_{\nu_\alpha}(E_\nu, \boldsymbol{f}_\Phi)
 =
 \frac{f_{\alpha,\oplus}^\pi}{2} 
 \frac{10^{\varphi(E_\nu, \boldsymbol{f}_\Phi)}}{E_\nu^2} \;.
\end{equation}
Among the flux models that we consider, a PCHIP has the greatest flexibility to reproduce various shapes of neutrino spectra, because it has no set shape, but instead molds itself into the one that best fits experimental observations, as opposed to the PL and PLC models, which have a predetermined shape.  It is closer to the PPL model, but has greater flexibility owing to using cubic splines.  Later (Sec.~\ref{sec:methods_stat}) we find that, indeed, a PCHIP offers comparative advantages over the other tree flux models that we consider to jointly reconstruct the UHE neutrino flux and cross section (Sec.~\ref{sec:results}).  Similarly to the PPL flux model, we use seven PCHIP nodes because this is about the minimum number needed to capture the main features of the neutrino spectrum.


\subsection{UHE $\nu N$ DIS cross section}
\label{sec:models-cross_section}

We model the UHE $\nu N$ DIS cross section based on the BGR18 calculation~\cite{Bertone:2018dse}, $d\sigma_0/dy$.  We allow for deviations from it inside each energy decade from $10^7$ to $10^{10}$~GeV, via the free shift parameters $f_{\sigma, 1}$, $f_{\sigma, 2}$, and $f_{\sigma, 3}$, \ie,
\begin{eqnarray}
 \label{equ:xsec}
 &&
 \frac{d\sigma(E_\nu, y, f_{\sigma, i})}{dy}
 = 
 \frac{d\sigma_0(E_\nu, y)}{dy}  
 \nonumber
 \\ &&
 \qquad\qquad
 \times
 \begin{cases}
  10^{f_{\sigma, 1}}, & 10^7 \leq E_\nu/{\rm GeV} < 10^8 \\
  10^{f_{\sigma, 2}}, & 10^8 \leq E_\nu/{\rm GeV} < 10^9 \\
  10^{f_{\sigma, 3}}, & 10^9 \leq E_\nu/{\rm GeV} \leq 10^{10} 
 \end{cases} \;.
\end{eqnarray}
When $f_{\sigma, 1} = f_{\sigma, 2} = f_{\sigma, 3} = 0$, \equ{xsec} matches the BGR18 calculation.  In our forecasts (Sec.~\ref{sec:methods_stat}), we find values for $\boldsymbol{f}_\sigma \equiv (f_{\sigma, 1}, f_{\sigma, 2}, f_{\sigma, 3})$ via fits to projected observations.  This parametrization of the UHE cross section is akin to the one used to measure the TeV--PeV cross section in \Refes~\cite{Bustamante:2017xuy, IceCube:2020rnc} and to forecast the measurement of the UHE cross section in \Refe~\cite{Esteban:2022uuw}.  

Previous works~\cite{Valera:2022ylt, Valera:2022wmu} based on the same event-rate calculation framework that we use here (Sec.~\ref{sec:event_rates}) parametrized instead the UHE cross section with a single floating parameter across $10^7$--$10^{10}$~GeV (see also \Refe~\cite{Denton:2020jft}), akin to what was used in the first measurement of the TeV--PeV cross section~\cite{IceCube:2017roe}.  There are two advantages to using multiple bins instead, as in \equ{xsec}.  First, it allows us not only to measure the energy dependence of the cross section, but also to find potential narrow nonstandard features in it, or to identify changes in its growth rate with energy that only turn on at a threshold energy; see \figu{main_result}.  Second, it ensures that the cross-section measurement uncertainty that we report for each energy decade reflects the statistical power of that decade, and is not artificially driven by the power of a different energy region that has a higher event rate. Figure~\ref{fig:main_result} illustrates this: the cross section is measured more precisely in the higher energy bins, where the event rate is higher, than in the lowest energy bin, where it is lower. 

In our analysis, a change in  the $\nu N$ DIS cross section affects the calculation of neutrino-induced event rates at two stages: during neutrino propagation inside the Earth and in the interaction of neutrinos inside the detector.  (During  propagation inside the Earth, we leave non-DIS sub-leading neutrino interactions~\cite{Garcia:2020jwr, NuPropEarth, Valera:2022ylt} unchanged.)  Using the parametrization of the cross section in \equ{xsec} makes computing neutrino propagation and detection more nuanced.  As neutrinos propagate underground, their energies are lowered via neutrino regeneration in NC interactions or, for $\nu_\tau$, in CC interactions.  As a result, for example, a neutrino that enters the Earth with an energy in the $10^9$--$10^{10}$~GeV decade and, therefore, interacts initially with a cross section scaled by $f_{\sigma, 3}$, could eventually lose enough energy to lie in the $10^8$--$10^9$~GeV decade, at which point it would interact instead with a cross section scaled by $f_{\sigma, 2}$.  Further energy losses might lower the  energy into the $10^7$--$10^8$~GeV decade, at which point the neutrino would interact with a cross section scaled by $f_{\sigma, 1}$.  

However, keeping track of how the relevant cross-section shift parameter changes during neutrino propagation inside the Earth would be a computationally demanding task within {\sc NuPropEarth} (see Sec.~\ref{sec:nu_detection-propagation}).  We deal with this in a simplified manner.  For given values of the scaling parameters, before starting to propagate a neutrino of a certain energy inside the Earth, we select what energy decade it falls in and pick out the cross-section scaling parameter associated to it, $f_{\sigma, 1}$, $f_{\sigma, 2}$, or $f_{\sigma, 3}$.  Then we propagate the neutrino inside the Earth assuming that the value of that shift parameter is common to all shift parameters.  The shift parameters are made equal only for the sake of speeding up the propagation, though.  At the end of propagation, the neutrino interacts inside the detector with a cross section scaled by $f_{\sigma, 1}$, $f_{\sigma, 2}$, or $f_{\sigma, 3}$---no longer taken to be equal---depending on what its final energy is.

The error we incur in by adopting the above simplification is small.  Since the average $\nu N$ DIS inelasticity at ultra-high energies is $\langle y \rangle \approx 0.25$, a neutrino retains about $75\%$ of its energy in each NC interaction it undergoes. This means that a neutrino would need to undergo eight consecutive NC interactions for its final energy to be one tenth of its original energy; this is as many interactions as it can undergo before falling into a lower energy decade.  Given that $\sigma_{\nu N}^{\rm NC} \approx \sigma_{\nu N}^{\rm CC}/3$~\cite{Bertone:2018dse}, the probability of a neutrino undergoing eight consecutive NC without disappearing due to a CC interaction is $(1/4)^8 \approx 0.002\%$.  Since for each neutrino flavor, energy, and direction we propagate $10^7$ neutrinos through the Earth, only about 150 of those would have propagated with an incorrect cross-section shift parameter.


\subsection{Atmospheric muon background}

We model the background of muon-induced events based on its baseline prescription (Sec.~\ref{sec:bg_rates}), but allowing for its rescaling via the free parameter $f_\mu$, \ie,
\begin{equation}
 \label{equ:diff_rate_mu_rescaled}
 \frac{dN^2_\mu}{dE_{\rm sh}^{\rm rec} d\theta_z^{\rm rec}}
 \rightarrow
 \frac{f_\mu}{\mathcal{N}_\mu}  
 \frac{dN^2_\mu}{dE_{\rm sh}^{\rm rec} d\theta_z^{\rm rec}} \;,
\end{equation}
where $\mathcal{N}_\mu$ is the baseline rate of all-sky, energy-integrated muons.  The parameter $f_\mu$ is the rescaled rate of muon-induced events; if $f_\mu = \mathcal{N}_\mu$, we recover the baseline prescription.  In our forecasts (Sec.~\ref{sec:methods_stat}) we find its value via fits to projected observations.

Like in \Refes~\cite{Valera:2022ylt, Valera:2022wmu}, we only allow for changes in the normalization of the atmospheric muon flux---including large ones, up to $f_\mu = 100$ (Table~\ref{tab:parameters})---but not in the shape of its energy spectrum.  Were the muon spectrum to extend to higher energies than in its baseline prescription, into the region where the UHE neutrino flux is expected to be larger, its influence in our forecasts might change.  Exploring that possibility requires dedicated studies beyond the scope of this paper.


\section{Statistical methods}
\label{sec:methods_stat}

\begingroup
\squeezetable
\begin{table*}[t!]
 \begin{ruledtabular}  
  \caption{\label{tab:parameters}\textbf{\textit{Free model parameters, their priors, and true values.}}  
 The flux parameters are different for each of the four flux fit models: PL, PLC, PPL, and PCHIP.  The cross-section parameters and the atmospheric muon parameter are the same regardless of the choice of flux fit model.  See Section~\ref{sec:models} for details.}
  \centering
  \renewcommand{\arraystretch}{1.4}
  \begin{tabular}{ccccccc}
   \multirow{2}{*}{\makecell{Flux fit\\model}} &
   \multicolumn{5}{c}{Free model parameter} 
   \\
   \cline{2-7} 
   &
   Symbol &
   Units &
   Description &
   Prior range\footnote{All the priors are uniform within their corresponding ranges.} &
   True value\footnote{We show true values only when they are available.  For the flux parameters, they refer to our benchmark UHE neutrino flux (Sec.~\ref{sec:nu_flux-benchmark}).} &
   Ref. \\
   \hline
   \multicolumn{7}{c}{Flux parameters, $\boldsymbol{f}_\Phi$} \\
   \hline 
   PL &
   $\Phi_0$ &
   GeV$^{-1}$~cm$^{-2}$~s$^{-1}$~sr$^{-1}$ &
   Flux norm.~at 10~PeV &
   [$10^{-27}$, $10^{-19}$] &
   $\cdots$ &
   \equ{flux_pl} \\
   &
   $\gamma$ &
   $\cdots$ &
   Spectral index &
   [0, 5] &
   $\cdots$ \vspace{0.1cm} &
   \\
   PLC &
   $\Phi_0$ &
   GeV$^{-1}$~cm$^{-2}$~s$^{-1}$~sr$^{-1}$ &
   Flux norm.~at 10~PeV &
   [$10^{-27}$, $10^{-19}$] &
   $\cdots$ &
   \equ{flux_plc} \\
   &
   $\gamma$ &
   $\cdots$ &
   Spectral index &
   [0, 5] &
   $\cdots$ &
   \\
   &
   $E_{\rm cut}$ &
   GeV &
   Cut-off energy &
   [$10^6$, $10^{11}$] &
   $\cdots$ &
   \vspace{0.1cm} \\
   PPL &
   $\Phi_1$ &
   GeV$^{-1}$~cm$^{-2}$~s$^{-1}$~sr$^{-1}$ &
   $E_\nu^{-2}$ flux norm., $10^{7}$--$10^{7.5}$~GeV &
   [$0$, $10^{-19}$] &
   $\cdots$ &
   \equ{flux_ppl} \\
   &
   $\Phi_2$ &
   GeV$^{-1}$~cm$^{-2}$~s$^{-1}$~sr$^{-1}$ &
   \qquad\qquad\qquad \ldots $10^{7.5}$--$10^{8}$~GeV &
   [$0$, $10^{-19}$] &
   $\cdots$ &
   \\
   &
   $\Phi_3$ &
   GeV$^{-1}$~cm$^{-2}$~s$^{-1}$~sr$^{-1}$ &
   \qquad\qquad\qquad \ldots $10^{8}$--$10^{8.5}$~GeV &
   [$0$, $10^{-19}$] &
   $\cdots$ &
   \\
   &
   $\Phi_4$ &
   GeV$^{-1}$~cm$^{-2}$~s$^{-1}$~sr$^{-1}$ &
   \qquad\qquad\qquad \ldots $10^{8.5}$--$10^{9}$~GeV &
   [$0$, $10^{-19}$] &
   $\cdots$ \\
   &
   $\Phi_5$ &
   GeV$^{-1}$~cm$^{-2}$~s$^{-1}$~sr$^{-1}$ &
   \qquad\qquad\qquad \ldots $10^{9}$--$10^{9.5}$~GeV &
   [$0$, $10^{-19}$] &
   $\cdots$ &
   \\
   &
   $\Phi_6$ &
   GeV$^{-1}$~cm$^{-2}$~s$^{-1}$~sr$^{-1}$ &
   \qquad\qquad\qquad \ldots $10^{9.5}$--$10^{10}$~GeV &
   [$0$, $10^{-19}$] &
   $\cdots$ &
   \vspace{0.1cm} \\
   PCHIP &
   $f_{\Phi, 1}$ &
   $\cdots$ &
   Log$_{10}(E_\nu^2 \Phi_\nu)$ at $10^7$~GeV\footnote{The units of $E_\nu^2 \Phi_\nu$ are $10^{-8}$~GeV~cm$^{-2}$~s$^{-1}$~sr$^{-1}$.  In practice, for speed-up, we centered the prior of each $f_{\Phi, i}$ ($i = 1, \ldots, 7)$ at its true value.  However, because the prior is flat and wide, this choice does not affect our results.} &
   [-13, -3] &
   -8.49 &
   \equ{flux_pchip} \\
   &
   $f_{\Phi, 2}$ &
   $\cdots$ &
   \qquad\qquad\quad \ldots~at $10^{7.5}$~GeV &
   [-13, -3] &
   -9.09 &
   \\
   &
   $f_{\Phi, 3}$ &
   $\cdots$ &
   \qquad\quad~~~ \ldots~at $10^{8}$~GeV &
   [-13, -3] &
   -8.55 &
   \\
   &
   $f_{\Phi, 4}$ &
   $\cdots$ &
   \qquad\qquad\quad \ldots~at $10^{8.5}$~GeV &
   [-13, -3] &
   -7.91 &
   \\
   &
   $f_{\Phi, 5}$ &
   $\cdots$ &
   \qquad\quad~~~ \ldots~at $10^{9}$~GeV &
   [-13, -3] &
   -7.61 &
   \\
   &
   $f_{\Phi, 6}$ &
   $\cdots$ &
   \qquad\qquad\quad \ldots~at $10^{9.5}$~GeV &
   [-13, -3] &
   -7.63 &
   \\
   &
   $f_{\Phi, 7}$ &
   $\cdots$ &
   \qquad\qquad~ \ldots~at $10^{10}$~GeV &
   [-13, -3] &
   -7.84 &
   \\
   \hline
   \multicolumn{7}{c}{Cross section parameters, $\boldsymbol{f}_\sigma$} \\
   \hline
   All
   &
   $f_{\sigma, 1}$ &
   $\cdots$ &
   $\sigma_{\nu N}$ shift, $10^7$--$10^8$~GeV   &
   [-2, 2] &
   0 &
   \equ{xsec} \\
   &
   $f_{\sigma, 2}$ &
   $\cdots$ &
   \qquad~~ \ldots~$10^8$--$10^9$~GeV &
   [-2, 2] &
   0 &
   \\
   &
   $f_{\sigma, 3}$ &
   $\cdots$ &
   \qquad~~~ \ldots~$10^9$--$10^{10}$~GeV &
   [-2, 2] &
   0 &
   \\
   \hline
   \multicolumn{7}{c}{Atmospheric muon background parameter, $f_\mu$ (nuisance)} \\
   \hline
   All &
   $f_\mu$ &
   $\cdots$ &
   Number atm.~$\mu$ &
   [0, 100] &
   $\cdots$ &
   \equ{diff_rate_mu_rescaled} \\
  \end{tabular}
 \end{ruledtabular}
\end{table*}
\endgroup

To produce our forecasts of joint measurement of the UHE neutrino spectrum and cross section,  we adopt a Bayesian approach based on generating and analyzing a large number of mock event samples that represent the expected response of the radio array of IceCube-Gen2.

We generate a mock observed event sample by assuming our benchmark model for the UHE neutrino flux (Sec.~\ref{sec:nu_flux-benchmark}) and the BGR18 $\nu N$ DIS cross section; we refer to them as the \textit{true flux} and \textit{true cross section} below and in \figu{main_result}.  Using the procedure described in Sec.~\ref{sec:nu_detection}, we compute the mock differential event rate of neutrino-initiated events, \equ{spectrum_tot}, to which we add the rate of events initiated by atmospheric muons (Sec.~\ref{sec:bg_rates}).  We interpret the differential event rate  as a joint probability distribution function in  $E_{\rm sh}^{\rm rec}$ and $\theta_z^{\rm rec}$, from which we randomly sample mock observed events.  Then we compare that observed event sample {\it vs.}~an event sample generated using test values of the flux and cross-section parameters, adopting for them the fit models described in Sec.~\ref{sec:models}.  We repeat this procedure many times, so as to average over all possible realizations of the observed event sample.

Below we describe the procedure step-by-step in detail to facilitate its independent implementation.  We carry it out separately for each choice of flux fit model---PL, PLC, PPL, and PCHIP.

\begin{enumerate}
 \item
  Taking our benchmark flux (Sec.~\ref{sec:nu_flux-benchmark}) as the true neutrino flux and the BGR18 $\nu N$ cross section (Sec.~\ref{sec:nu_detection-propagation}) as the true cross section, \ie, $f_{\sigma, 1} = f_{\sigma, 2} = f_{\sigma, 3} = 0$ in \equ{xsec}, compute the differential rate of observed neutrino-induced events, \equ{spectrum_tot}, and the baseline rate of muon-induced events, \equ{diff_rate_mu_rescaled} with $f_\mu = \mathcal{N}_\mu$, and, with them, the total differential observed event rate,
  \begin{equation}
   \frac{d^2N_{\rm sh}}{dE_{\rm sh}^{\rm rec} d\theta_z^{\rm rec}}
   =
   \frac{d^2N_{\nu}}{dE_{\rm sh}^{\rm rec} d\theta_z^{\rm rec}}
   +
   \frac{d^2N_{\mu}}{dE_{\rm sh}^{\rm rec} d\theta_z^{\rm rec}} \;.
  \end{equation}
  (Figure~\ref{fig:event_rate} shows these distributions, though only summed over all energies or over all directions.)  Later steps in the calculation (2--11) contrast samples of observed events drawn from this distribution against test samples generated using test choices of the flux and cross-section parameters.
 \item
  Compute the probability distribution function of observed events,
  \begin{equation}
   \label{equ:prob_dist_mock}
   \qquad\qquad
   \mathcal{P}(E_{\rm sh}^{\rm rec}, \theta_z^{\rm rec})
   =
   \frac{1}{\mathcal{N}_{\rm obs}}
   \frac{d^2N_{\rm sh}
   (E_{\rm sh}^{\rm rec}, 
   \theta_z^{\rm rec})}{dE_{\rm sh}^{\rm rec} d\theta_z^{\rm rec}} \;,
  \end{equation}
  where $\mathcal{N}_{\rm obs} = \mathcal{N}_\nu + \mathcal{N}_\mu$ is the all-sky, energy-integrated mean number of observed events.  For our choice of true neutrino flux, $\mathcal{N}_\nu = 33.1$ neutrino-initiated events per year (Sec.~\ref{sec:event_rates}).  For the baseline computation of the atmospheric muon background, $\mathcal{N}_\mu = 0.054$ muon-initiated events per year (Sec.~\ref{sec:bg_rates}).
  \item
   Randomly sample the number of observed events, $N_{\rm obs}$, from a Poisson distribution whose central value is equal to the mean expectation, $\mathcal{N}_{\rm obs}$.  Then generate a sample of $N_{\rm obs}$ detected events, $\left\{ e_i \right\}_{i=1}^{N_{\rm obs}}$, each consisting of a pair of reconstructed energy and direction, $e_i \equiv (E_{{\rm sh}, i}^{\rm rec}, \theta_{z, i}^{\rm rec})$, whose values are randomly sampled from \equ{prob_dist_mock}.
 \item
  For the chosen flux fit model---PL, PLC, PPL, or PCHIP---generate a random variate of the model parameters, $\boldsymbol{\theta} \equiv (\boldsymbol{f}_\Phi, \boldsymbol{f}_\sigma, f_\mu)$ (Sec.~\ref{sec:models}).  (In practice, we sample the parameter values from prior distributions; more on this later.)
 \item
  Compute the isotropic flux of $\nu_\alpha$ and $\bar{\nu}_\alpha$ at the surface of the Earth, $\Phi_{\nu_\alpha}(E_\nu, \boldsymbol{f}_\Phi) = \Phi_{\bar{\nu}_\alpha}(E_\nu, \boldsymbol{f}_\Phi)$.
 \item
  Propagate separately $\nu_e$, $\bar{\nu}_e$, $\nu_\mu$, $\bar{\nu}_\mu$, $\nu_\tau$, and $\bar{\nu}_\tau$ from the surface of the Earth to IceCube-Gen2 (Sec.~\ref{sec:nu_detection}), using the $\nu N$ DIS cross section modified by the parameters $\boldsymbol{f}_\sigma$, \equ{xsec}.  The resulting fluxes at the detector are no longer isotropic, \ie, $\Phi_{\nu_\alpha}^{\rm det}(E_\nu, \cos \theta_z, \boldsymbol{f}_\Phi, \boldsymbol{f}_\sigma)$, and similarly for $\bar{\nu}_\alpha$.
 \item
  Compute the corresponding differential rate of neutrino-induced events in the radio array of IceCube-Gen2 (Sec.~\ref{sec:nu_detection}), $d^2N_\nu (\boldsymbol{f}_\Phi, \boldsymbol{f}_\sigma) / dE_{\rm sh}^{\rm rec} d\theta_z^{\rm rec}$, \equ{spectrum_tot}.  In doing so, the parameters $\boldsymbol{f}_\sigma$ also modify the $\nu N$ DIS cross section used at detection.
 \item
  Compute the differential rate of events induced by atmospheric muons, re-scaled by the factor $f_\mu$, $d^2N_\mu (f_\mu) / dE_{\rm sh}^{\rm rec} d\theta_z^{\rm rec}$, \equ{diff_rate_mu_rescaled}.
 \item
  Integrate the differential event rates to find the all-sky, energy-integrated number of events due to neutrinos, $N_\nu(\boldsymbol{f}_\Phi, \boldsymbol{f}_\sigma)$, and  muons, $N_\mu(f_\mu) = f_\mu$.  The total number of events is $N(\boldsymbol{\theta}) \equiv N_\nu(\boldsymbol{f}_\Phi, \boldsymbol{f}_\sigma) + f_\mu$.
 \item
  Compute the fraction of events in the sample that is due to neutrinos (\ie, the signal), $\mathcal{F}_\nu(\boldsymbol{\theta}) \equiv N_\nu(\boldsymbol{f}_\Phi, \boldsymbol{f}_\sigma) / N(\boldsymbol{\theta})$, and the fraction that is due to muons (\ie, the background), $\mathcal{F}_\mu(\boldsymbol{\theta}) \equiv f_\mu / N(\boldsymbol{\theta})$.
 \item
  For the $i$-th event in the sample, $e_i$, compute the partial likelihood
  \begin{align}
   \label{equ:likelihood_partial}
   & \qquad
   \mathcal{L}_i(\boldsymbol{\theta})
   =
   \mathcal{F}_\nu(\boldsymbol{\theta}) p_\nu(e_i \vert \boldsymbol{f}_\Phi, \boldsymbol{f}_\sigma)
   +
   \mathcal{F}_\mu(\boldsymbol{\theta}) p_\mu(e_i \vert f_\mu) \;,
  \end{align}
  where the probability density of this event being due to a neutrino is
  \begin{align}
   \label{eq:nu_pdf}
   & \quad
   p_\nu(e_i \vert \boldsymbol{f}_\Phi, \boldsymbol{f}_\sigma)
   =
   \frac{1}{N_\nu (\boldsymbol{f}_\Phi, \boldsymbol{f}_\sigma)}
   \nonumber \\
   & \qquad\qquad\qquad\qquad
   \times \left.
   \frac{d^2N_\nu (\boldsymbol{f}_\Phi, \boldsymbol{f}_\sigma)}{dE_{\rm sh}^{\rm rec} d\theta_z^{\rm rec}} 
   \right\vert_{E_{{\rm sh}, i}^{\rm rec}, \theta_{z, i}^{\rm rec}}
   \;,
  \end{align}
  and the probability density of it being due to a muon is
  \begin{align}
   \label{eq:mu_pdf}
   & \quad
   p_\mu(e_i \vert f_\mu)
   =
   \frac{1}{f_\mu}
   \left.
   \frac{d^2N_\mu (f_\mu)}{dE_{\rm sh}^{\rm rec} d\theta_z^{\rm rec}} 
   \right\vert_{E_{{\rm sh}, i}^{\rm rec}, \theta_{z, i}^{\rm rec}}
   \;.
  \end{align}
  Compute \equ{likelihood_partial} for each of the $N_{\rm obs}$ events in the sample.
 \item
  Compute the unbinned extended Poisson likelihood for the full event sample,
  \begin{align}
   \label{equ:likelihood}
   & \qquad
   \mathcal{L}\left(\left\{e_i\right\}_{i=1}^{N_{\rm obs}} | \boldsymbol{\theta}\right)
   =
   \frac{e^{-N(\boldsymbol{\theta})} N(\boldsymbol{\theta})^{N_{\rm obs}}}{N(\boldsymbol{\theta})!} 
   \prod_{i=1}^{N_{\rm obs}} \mathcal{L}_i(\boldsymbol{\theta}) \;.
  \end{align}
  We use an unbinned likelihood to avoid our results depending on our choice of bin size.
 \item
  Using Bayes' theorem, compute the corresponding posterior probability density,
  \begin{align}
   \label{equ:posterior}
   &
   \mathcal{P}\left(\left\{e_i\right\}_{i=1}^{N_{\rm obs}} | \boldsymbol{\theta}\right)
   =
   \frac{\mathcal{L}\left(\left\{e_i\right\}_{i=1}^{N_{\rm obs}} | \boldsymbol{\theta}\right) \pi(\boldsymbol{\theta})}{\mathcal{Z}\left(\left\{e_i\right\}_{i=1}^{N_{\rm obs}}\right)} \;,
  \end{align}
  where $\pi (\boldsymbol{\theta}) \equiv \pi(\boldsymbol{f}_\Phi) \pi(\boldsymbol{f}_\sigma) \pi(f_\mu)$ is the prior distribution on the model parameters.  For each parameter, we use a wide, uniform prior, and we assume no correlations between them; see Table~\ref{tab:parameters}.  The denominator in \equ{posterior} is the model evidence, and is obtained by integrating the numerator over the whole model parameter space of $\boldsymbol{\theta}$.  We compute it using \textsc{UltraNest}~\cite{Ultranest}, an efficient importance nested sampler~\cite{Buchner:2014, Buchner:2017}.
 \item
  Keeping the same observed event sample, repeat steps 4--13 for many different random variates $\boldsymbol{\theta}$, using {\sc UltraNest}, until the parameter space has been thoroughly explored, and the posterior has been evaluated throughout it.
 \item
  Repeat steps 3--14, for $10^4$ random observed event samples.  After that, compute the posterior averaged over all the realizations of observed event samples, $\bar{\mathcal{P}}(\boldsymbol{\theta})$, which we use to make our forecasts.  We maximize it to compute the best-fit values of the model parameters, and we integrate it to find their credible intervals.
\end{enumerate}


\section{Results}
\label{sec:results}

\begin{figure*}[t]
    \centering
    \includegraphics[width=\textwidth]{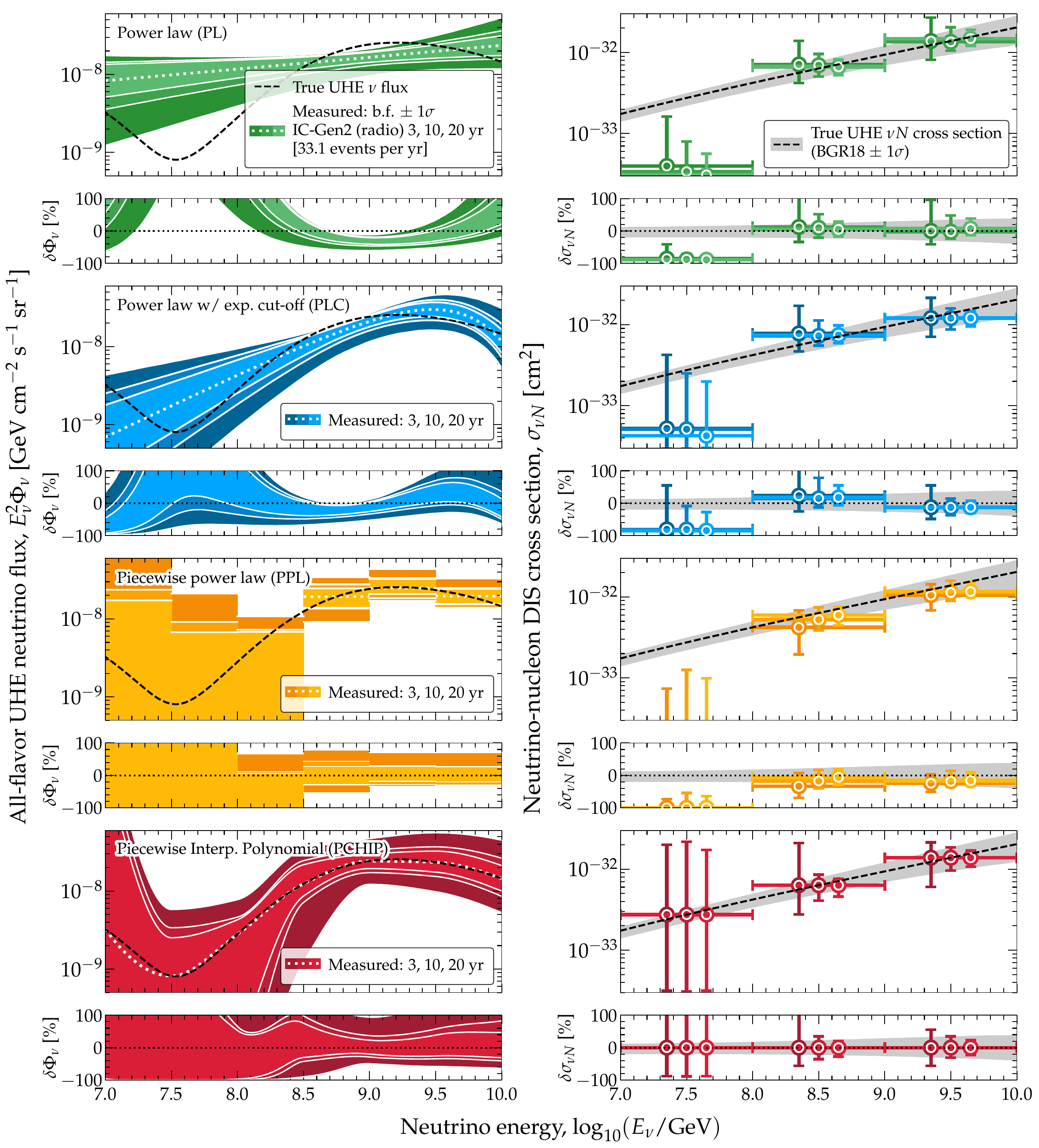} 
    \caption{\textbf{\textit{Forecasts of the joint measurement of the UHE neutrino energy spectrum and the neutrino-nucleon cross section.}}  Each row shows the performance of using one of our models of the neutrino spectrum (Sec.~\ref{sec:models-flux}) in a fit to simulated observed event samples in the radio array of IceCube-Gen2 (Sec.~\ref{sec:nu_detection}), after 3, 10, and 20 years of exposure.  \textit{Left column:}  Measured neutrino energy spectrum.  The true UHE neutrino flux is our benchmark flux, representative of theoretical predictions (Sec.~\ref{sec:nu_flux}). \textit{Right column:} Measured neutrino-nucleon cross section, reconstructed in three decade-wide energy bins.  The true cross section is the BGR18 model~\cite{Bertone:2018dse}.  \textit{The PL and PLC flux models yield high precision---due to their having a small number of free parameters---but low accuracy---due to their rigid shape.  The PPL and PCHIP flux models yield comparable precision but superior accuracy---due to their flexible shape.} See Sec.~\ref{sec:results} for details.
    }
    \label{fig:alternative_models}
\end{figure*}


\subsection{General trends}

Figure~\ref{fig:alternative_models}  (also \figu{main_result}) illustrates our forecasts of the joint measurement of the UHE neutrino spectrum and $\nu N$ DIS cross section in the radio array of IceCube-Gen2.   We show results assuming each of our four neutrino spectrum fit models in turn (Sec.~\ref{sec:models-flux})---PL, PLC, PPL, and PCHIP---and adopting for the true UHE neutrino flux our benchmark flux from Sec.~\ref{sec:nu_flux-benchmark}.  To produce our main forecasts, we assume baseline energy and angular resolution of $\sigma_\epsilon = 0.1$ and $\sigma_{\theta_z} = 2^\circ$, respectively (Sec.~\ref{sec:nu_detection}).  Later, we explore the impact of alternative choices.

There are common trends shared by the four sets of results.  At intermediate energies, from 100~PeV to 1~EeV, the measurements are most precise, though not necessarily accurate---except for PCHIP, for which results are always accurate.   At low and high energies, the measurements worsen due to low event rates.  At low energies, in the tens of PeV, this is due to the small effective volume of the radio array of IceCube-Gen2, since Askaryan emission weakens~\cite{Schroder:2016hrv}.  This stresses the need for complementary measurements at tens of PeV, \eg, by the optical component of IceCube-Gen2~\cite{IceCube-Gen2:2020qha}, TAMBO~\cite{Romero-Wolf:2020pzh, Thompson:2023mob}, or Trinity~\cite{Otte:2018uxj, Brown:2021lef}.  At high energies, above 1~EeV, the measurements worsen because our benchmark flux model decreases, which is representative of flux predictions.  Longer detector exposure improves the accuracy of the measurements, and more slowly, their precision; we show results for 3, 10, and 20 years.


\subsection{Power law (PL)}
\label{sec:results-pl}

Figure~\ref{fig:alternative_models} shows that, as expected, the PL flux model is too simple and rigid to capture the features of our benchmark flux, \ie, the dip around $10^{7.5}$~GeV and the bump around $10^9$~GeV, even after 20 years of exposure.  After 10 or 20 years, the spectrum is reconstructed with high precision---due to the low number of flux model parameters---but with atrocious accuracy: the allowed flux band overestimates the true neutrino flux at low energies by about one order of magnitude and at high energies by about 100\%, and underestimates it by tens of percent at intermediate energies.  

This, in turn, affects the accuracy of the joint measurement of the cross section.  Because of the partial degeneracy between flux and cross section in the computation of event rates (Sec.~\ref{sec:synopsis-measuring}), when the measured flux is too high relative to its true value, the measured cross section is too low relative to its own. 
Figure~\ref{fig:alternative_models} shows that this trade-off is flagrant in the lowest energy decade, where the measured cross section is offset from its true value by about 100\%, regardless of the exposure time.  In the intermediate and high energy decades, the cross section can be measured within tens of percent after 10--20 years.

Given that all but the barest predictions posit UHE neutrino spectra with shapes more complex than a power law (see Fig.~2 in \Refe~\cite{Valera:2022wmu}), using the PL model in the fits would knowingly run a high risk of misreconstructing the flux and cross section. \textbf{\textit{Lacking prior knowledge of the true shape of the UHE neutrino spectrum, we recommend against using the PL model.}}


\subsection{Power law with exponential cut-off (PLC)}
\label{sec:results-plc}

Figure~\ref{fig:alternative_models} shows that the PLC flux model performs marginally better than the PL model, but retains its main shortcomings.  Like with the PL model, the flux is measured precisely---less so due to having one more parameter than the PL model---but inaccurately.  At low energies, like for the PL model, flux reconstruction is particularly inaccurate because the event rate drops.  At intermediate and high energies, it is more accurate than for the PL model because the exponential cut-off in the PLC model makes it possible to fit the bump-like spectrum of the true flux.  Regarding the cross section, the PLC model performs similarly to the PL model: it undershoots the cross section at the lowest energies, and measures it to within tens of percent at intermediate and high energies, centered close to its true value.

Were the true neutrino spectrum more closely a power law with a bump-like feature on top of it, like in the prediction of \Refe~\cite{Fang:2017zjf} (flux model 8 in \Refe~\cite{Valera:2022wmu}), then the PLC flux model would perform better.  \textbf{\textit{Like for the PL model, lacking prior knowledge of the true shape of the energy spectrum, we recommend against using the PLC model.  Yet, if pressed to use a flux model with a low number of free parameters, PLC is preferable to PL.}} 


\subsection{Piecewise power law (PPL)}
\label{sec:results-ppl}

Figure~\ref{fig:alternative_models} shows that the PPL flux model is able to reconstruct the shape of our benchmark neutrino spectrum with a precision comparable to that of the PL and PLC models at intermediate and high energies, but with higher accuracy.  At low energies, where our benchmark flux is low, the model allows only to place upper limits on it.  Regarding the cross section, the PPL model also roughly matches the precision of the PL and PLC models in the intermediate and high energy decades, but undershoots the true cross section in the lowest energy decade by about 100\%.

Unlike the PL and PLC models, and similarly to the PCHIP model below, the PPL model does not impose a fixed shape on the neutrino energy spectrum across the entire energy range.  The PPL model has the flexibility to reconstruct diverse shapes of the neutrino spectrum.  The width of each $E_\nu^{-2}$ energy segment in the PPL prescription, \equ{flux_ppl}, limits how closely the true shape of the energy spectrum can be approximated, and the precision with which the flux can be measured.  Using more PPL energy segments in the PPL would allow for a finer reconstruction of the shape of the energy spectrum, but could also worsen the measurement precision by introducing more free model parameters to fit. 

\textbf{\textit{Lacking prior knowledge of the true shape of the neutrino energy spectrum, we recommend using the PPL model to ensure sensitivity to the large variety of possible spectrum shapes.}}  However, in energy ranges with low event rates, PPL might only set upper limits on the flux and might report values of the cross section significantly offset from the real ones.  


\subsection{Piecewise Cubic Hermite Interpolating Polynomial (PCHIP)}
\label{sec:results-pchip}

Figure~\ref{fig:alternative_models} shows that the PCHIP flux fit model outperforms the other models in accuracy, at all energies, while approximating their precision at intermediate and high energies.  The superiority of PCHIP rests not from achieving higher precision, but from guaranteeing accuracy: the best-fit flux and cross section measured using PCHIP are always centered on their true values, even at low exposure times.  

The neutrino spectrum is measured accurately across the full energy range.  For a low exposure of 3 years, the measurement precision is worse than that of the other flux fit models, on account of the larger number of model parameters of the PCHIP model.  But, after 10--20 years, the precision becomes comparable to that of the other models at intermediate and high energies, reaching 50\%--25\%.  At low energies, it is worse due to the paucity of events, comparable to that of the PPL model.

The cross section is also measured accurately across all energy decades.  In the intermediate and high energy decade, after 10--20 years, the measurement precision is comparable to that of the other flux fit models.  In the low energy decade, the precision is significantly worse; there, similarly to the PPL model, the lack of a rigid spectrum shape means that the few events that are available are insufficient to make a precise measurement.

\textbf{\textit{Given the guaranteed accuracy of the PCHIP flux fit model, and its precision on par with other models, we recommend using it, especially if at least a few tens of events are available.}}


\begin{figure}[t]
 \centering
 \includegraphics[width=\columnwidth]{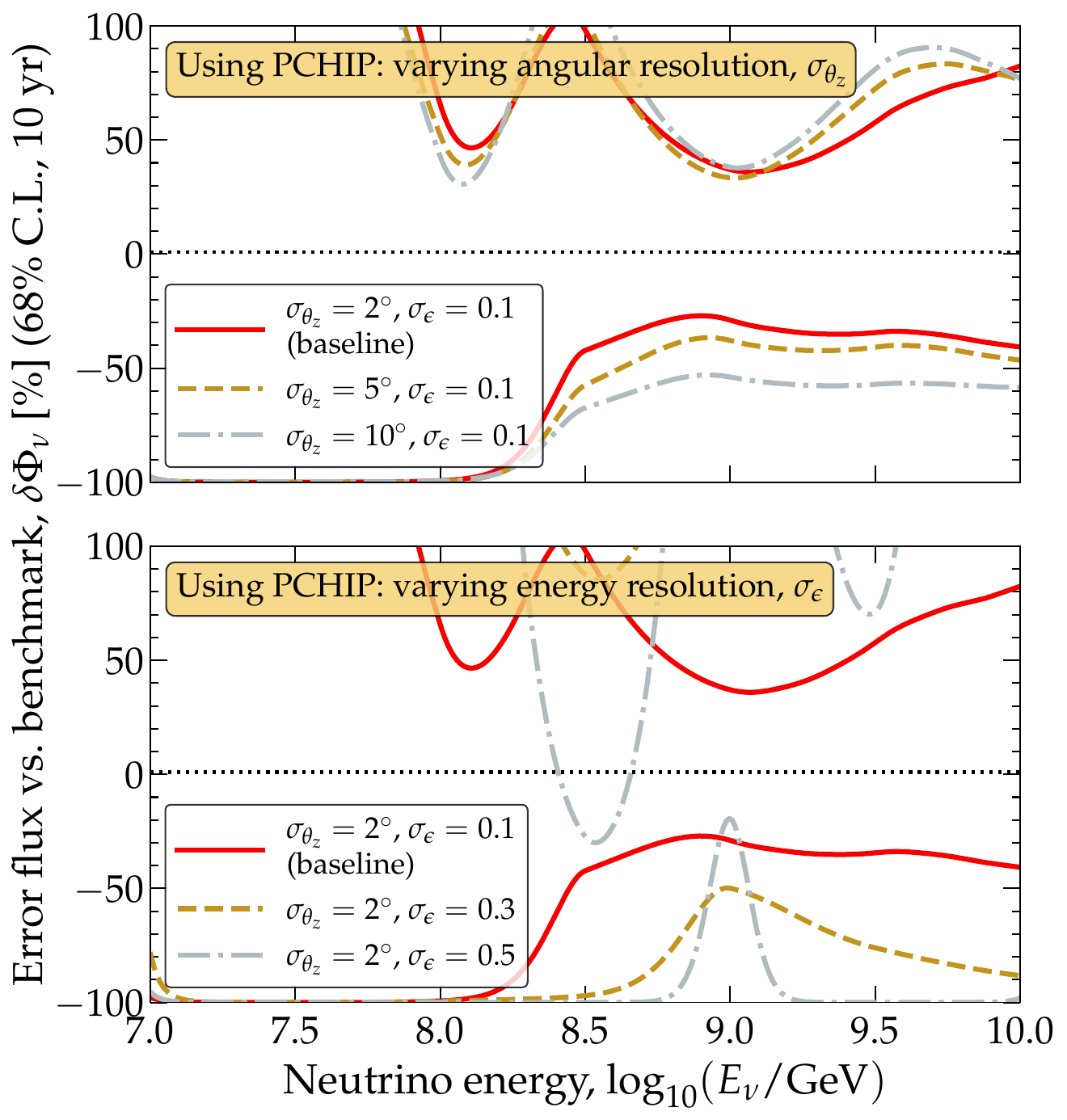}
 \caption{\textbf{\textit{Relative error in the measurement of the UHE neutrino flux, for varying detector resolution.}}  The detector is the radio array of IceCube-Gen2.  The comparison is between the flux measured using the PCHIP flux fit model, \equ{flux_pchip}, and our benchmark neutrino flux (Sec.~\ref{sec:nu_flux-benchmark}).  \textit{Top:} Varying the detector angular resolution.  \textit{Bottom:} Varying the detector energy resolution.  See Sec.~\ref{sec:event_rates} for how the detector resolution affects the computation of event rates, and Sec.~\ref{sec:results-resolution} for details.}
 \label{fig:res_flux}
\end{figure}

\begin{figure}[t]
 \centering
 \includegraphics[width=\columnwidth]{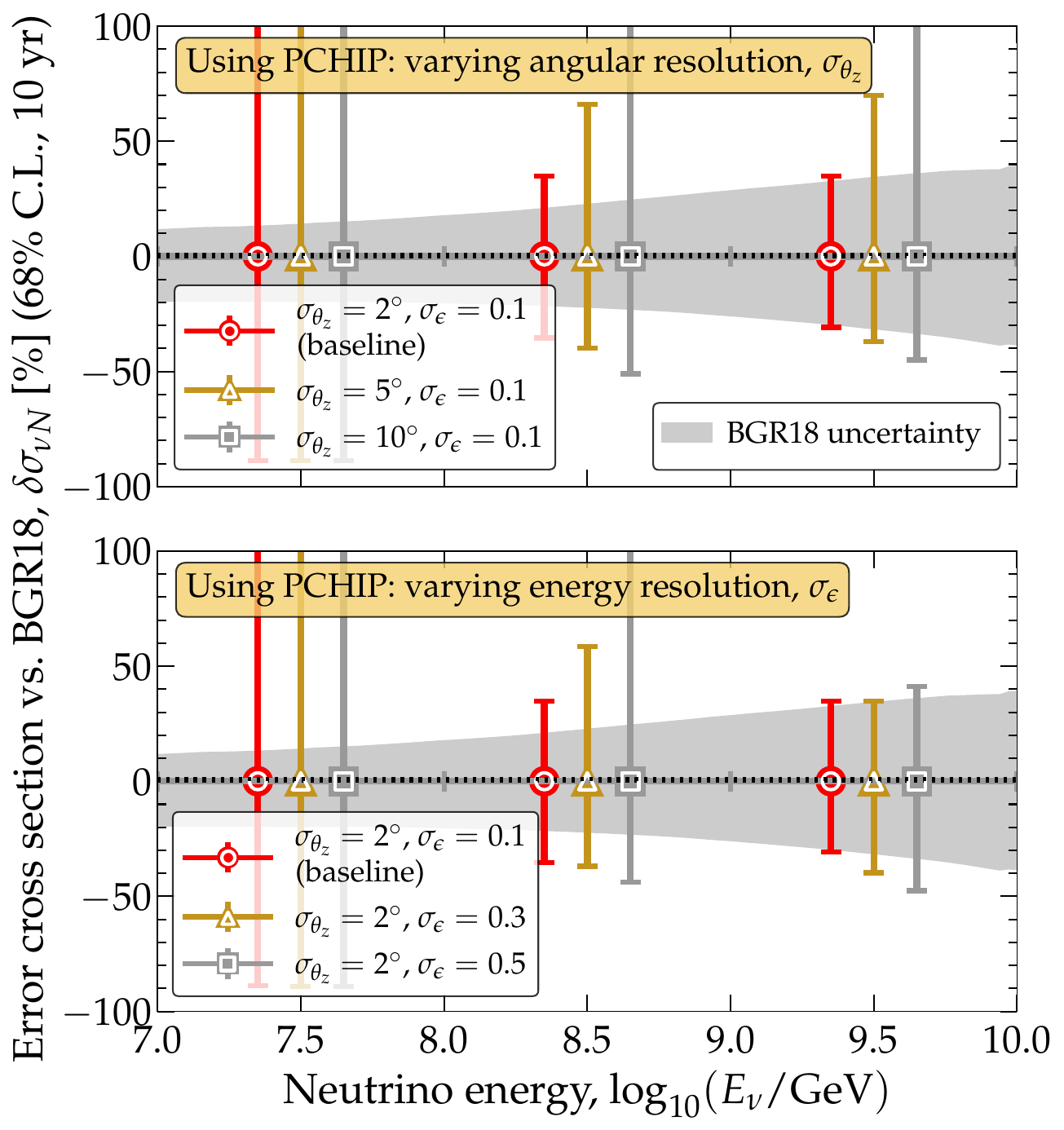}
 \vspace*{-0.5cm}
 \caption{\textbf{\textit{Relative error in the measurement of the UHE $\nu N$ DIS cross section, for varying detector resolution.}}  Same as \figu{res_flux}, but the comparison is between the measured cross section, \equ{xsec}, (assuming the PCHIP flux fit model, \equ{flux_pchip}) and the BGR18 calculation of the cross section~\cite{Bertone:2018dse}.  \textit{Top:} Varying the detector angular resolution.  \textit{Bottom:} Varying the detector energy resolution.  See Sec.~\ref{sec:results-resolution} for details.}
 \label{fig:res_xsec}
\end{figure}

\subsection{Impact of the energy and angular resolution}
\label{sec:results-resolution}

In Secs.~\ref{sec:results-pl}--\ref{sec:results-pchip}, we generated our main results using the baselines values of $\sigma_{\theta_z} = 2^\circ$ and $\sigma_\epsilon = 0.1$ for the angular and energy resolution of the detector (Sec.~\ref{sec:nu_detection}), the same ones used in related forecasts~\cite{Valera:2022ylt, Valera:2022wmu} based on similar techniques as here. However, the capabilities of upcoming UHE neutrino telescopes are still under development, including those of IceCube-Gen2, so we explore the effect on our results of changing the detector resolution, focusing on the PCHIP flux model.

Figure~\ref{fig:res_flux} shows the effect of changing the detector resolution on the measurement of the neutrino spectrum.  Poorer angular resolution worsens the precision moderately, but poorer energy resolution, even one slightly poorer than the baseline, significantly distorts the reconstructed energy spectrum.  This is because it becomes harder to infer the energy spectrum when features in the energy distribution are washed out by poor energy resolution.  This sets a loose design target of $\sigma_\epsilon \approx 0.1$ (see Sec.~\ref{sec:bg_rates} for a definition) for the detector energy resolution needed in order to measure the UHE neutrino spectrum using a flexible parametrization such as PCHIP.

Figure~\ref{fig:res_xsec} shows the effect of changing the detector resolution on the measurement of the cross section. Poorer angular resolution worsens the precision appreciably across all energy decades.  This is especially evident in the results for $\sigma_{\theta_z} = 10^\circ$.  Given that the measurement of the cross section stems from comparing the attenuation of the neutrino flux from different near-horizontal directions, about $\pm 5^\circ$ around the horizon~\cite{Valera:2022ylt}, a resolution of $10^\circ$ dramatically reduces the measurement precision.  

The impact of poorer energy resolution on the cross section is more nuanced.  In the lowest energy decade, its impact does not fundamentally change the outcome obtained using the baseline resolution, since measurements are already limited by low event rates.  In the intermediate and high energy decades, where the event rate is higher, the impact of poorer angular resolution is more evident.  The intermediate energy decade is more affected by it as a result of the misreconstruction of the energy spectrum, as shown in \figu{res_flux}.  This sets a loose design target of $\sigma_{\theta_z} \approx 2^\circ$ and $\sigma_{\epsilon} \approx 0.1$ in order to measure the energy dependence of the cross section.

Conveniently, the loose design targets for the detector resolution that we have found above---around our baseline choices of $\sigma_\epsilon = 0.1$ and $\sigma_{\theta_z} = 2^\circ$---are the same ones that would enable in the radio array of IceCube-Gen2, and other UHE neutrino telescopes, the discovery of the diffuse flux of UHE neutrinos and distinguishing between competing flux predictions~\cite{Valera:2022wmu}, measuring the UHE $\nu N$ cross section normalization~\cite{Valera:2022ylt}, finding point sources of UHE neutrinos~\cite{Fiorillo:2022ijt}, and looking for the decay of heavy dark matter into UHE neutrinos~\cite{Fiorillo:2023clw}.


\section{Summary and outlook}
\label{sec:summary}

The discovery of ultra-high-energy (UHE) neutrinos, with energies in excess of 100~PeV, would bring  insight into long-standing open questions in astrophysics and particle physics.  In preparation for the near-future discovery opportunities brought about by upcoming UHE neutrino telescopes, we have introduced methods to jointly measure two essential, but so-far-unknown quantities: the UHE neutrino flux and the neutrino-nucleon cross section, including their dependence with neutrino energy, and without prior knowledge of either.  Achieving this would unlock in earnest the potential of UHE neutrino telescopes to perform measurements unhampered, inasmuch as possible, by preconceptions of the size and shape of the flux and cross section that, especially for the former, are laden with large uncertainty from theory.

Our methods are of general applicability; we have presented them at length to facilitate their implementation.  We have illustrated them via forecasts of the capabilities of the planned radio array of the IceCube-Gen2 neutrino telescope, based on state-of-the-art simulations of it. 

In light of the large variety of shape and size in the predictions of the UHE neutrino flux, we have focused on flexible parametrizations of the neutrino spectrum that can capture this variety, aiming for measurement precision and accuracy, but favoring the latter over the former.  We explored four analysis models for the shape of the neutrino spectrum: two with rigid shape---a power law (PL) and a power law with an exponential cut-off (PLC)---and two with flexible, adaptable shape---a piecewise power law (PPL) and a Piecewise Cubic Hermite Interpolating Polynomial (PCHIP).

The PL and PLC flux models, while appealing in concept due to their simplicity, are unable to capture features of the neutrino spectrum---dips and bumps, stemming from the neutrino production processes---that are commonplace in theory predictions.  They achieve relatively high precision but atrocious accuracy, both in measuring the neutrino spectrum and the cross section.  This underscores the necessity for more nuanced and adaptable flux models.

The PPL and PCHIP flux models provide the required flexibility, though at the cost of introducing more free model parameters.  PPL stands out for its inherent flexibility, capable of accommodating a diverse range of spectrum shapes, though its accuracy is moderated by the challenges posed by low event rates.  PCHIP affords yet greater flexibility, but with unwavering accuracy, even when confronted with low event rates.  To use flexible flux models, like PCHIP, the detector requires a resolution of about 10\% per energy decade in the energy of detected events and about $2^\circ$ in the direction of detected events.  Conveniently, these loose design targets roughly match the projected performance of the radio array of IceCube-Gen2 presently under study.

Access to a new energy regime motivates revisiting analysis choices.  We present our methods and forecasts of the joint measurement of the UHE neutrino spectrum and neutrino-nucleon cross section with the goal of exploiting the full potential of UHE neutrino telescopes.

\medskip


\section*{Acknowledgements}

VBV would like to thank the Instituto de F\'isica Corpuscular (IFIC), Universidad de Valencia, for their hospitality during part of the development of this work.  MB and VBV are supported by {\sc Villum Fonden} under project no.~29388.  This work used resources provided by the High-Performance Computing Center at the University of Copenhagen. This work has been partially supported by the MCIN/AEI/10.13039/501100011033 of Spain under grant PID2020-113644GB-I00 and by the European Union’s Framework Programme for Research and Innovation Horizon 2020 (2014–2020) under grant H2020-MSCA-ITN-2019/860881-HIDDeN.

\twocolumngrid
\bibliography{refs.bib}


\end{document}